\documentclass[useAMS,usenatbib,fleqn]{mn2e}
\usepackage{graphicx,color,pdfpages,amsmath,float}
\newcommand{\Mpch}{\mbox{ $h^{-1}$ Mpc}}

\newcommand{\be}{\begin{equation}}
\newcommand{\ee}{\end{equation}}

\newcommand{\erfc}{\textrm{erfc}}

\def\ltsima{$\; \buildrel < \over \sim \;$}
\def\simlt{\lower.5ex\hbox{\ltsima}}
\def\gtsima{$\; \buildrel > \over \sim \;$}
\def\simgt{\lower.5ex\hbox{\gtsima}}
\def\la{$\langle$}

\title{The Mass-Concentration-Redshift Relation of Cold and Warm Dark Matter Halos}

\author[Ludlow et al.] {\parbox{18cm}{
    Aaron D. Ludlow$^{1,\star}$,
    Sownak Bose$^{1}$,
    Ra\'ul E. Angulo$^{2}$,
    Lan Wang$^{3}$,
    Wojciech A. Hellwing$^{1,4,5}$,
    Julio F. Navarro$^{6}$,
    Shaun Cole$^{1}$,
    Carlos S. Frenk$^{1}$
  }\vspace{0.3cm}\\
$^{1}${Institute for Computational Cosmology, Dept. of Physics, Univ. of
  Durham, South Road, Durham  DH1 3LE, UK}\\
$^{2}${Centro de Estudios de F\'isica del Cosmos de Arag\'on,  
  Plaza San Juan 1, Planta-2, 44001, Teruel, Spain}\\
$^{3}${Key Laboratory of Computational Astrophysics, National
  Astronomical Observatories, Chinese Academy of Sciences,} \\
  {~20A Datun Road, Chaoyang District, Beijing, China}\\
$^{4}${Institute of Cosmology and Gravitation, University of
  Portsmouth, Dennis Sciama Building, Portsmouth P01 3FX, UK}\\
$^{5}${Interdisciplinary Centre for Mathematical and Computational
  Modelling (ICM), University of Warsaw, ul. Pawi\'nskiego 5a, Warsaw, Poland}\\
$^{6}${Senior CIfAR Fellow. Department of Physics \& Astronomy, University of
  Victoria, Victoria, BC, V8P 5C2, Canada}\\
}

\begin{document}

\maketitle 

\begin{abstract}
  We use a suite of cosmological simulations to study the mass-concentration-redshift 
  relation, $c({\rm M},z)$, of dark matter halos. Our simulations include standard 
  $\Lambda$-cold dark matter (CDM) models, and additional runs with truncated power 
  spectra, consistent with a thermal warm dark matter (WDM) scenario. We find that 
  the mass profiles of CDM and WDM halos are self-similar and well approximated by 
  the Einasto profile. The $c({\rm M},z)$ relation of CDM halos is monotonic: 
  concentrations decrease with increasing virial mass at fixed redshift, and decrease 
  with increasing redshift at fixed mass. The mass accretion histories (MAHs) of CDM 
  halos are also scale-free, and can be used to infer concentrations directly. These 
  results do not apply to WDM halos: their MAHs are not scale-free because of the 
  characteristic scale imposed by the power-spectrum suppression. Further, the WDM 
  $c({\rm M},z)$ relation is non-monotonic: concentrations peak at a mass scale 
  dictated by the truncation scale, and decrease at higher and lower masses. We 
  show that the assembly history of a halo can still be used to infer its 
  concentration, provided that the total mass of its progenitors is considered 
  (the ``collapsed mass history''; CMH), rather than just that of its main ancestor. 
  This exploits the scale-free nature of CMHs to derive a simple scaling that 
  reproduces the mass-concentration-redshift relation of both CDM and WDM halos over 
  a vast range of halo masses and redshifts. Our model therefore provides a robust 
  account of the mass, redshift, cosmology and power spectrum dependence of dark 
  matter halo concentrations.
\end{abstract}

\renewcommand{\thefootnote}{\fnsymbol{footnote}}
\footnotetext[1]{E-mail: aaron.ludlow@durham.ac.uk} 

\section{Introduction}
\label{SecIntro}

It is now well established that 
the density profiles of equilibrium dark matter (DM) halos have an
an approximately universal form that can be reproduced 
by rescaling a simple formula \citep[hereafter NFW]{Navarro1996,Navarro1997},
\begin{equation}
  \frac{\rho(r)}{\rho_{\rm{crit}}}=\frac{\delta_c}{(r/r_s)(1+r/r_s)^2}.
  \label{eq:rho_nfw}
\end{equation}
Known as the ``NFW profile'', eq.~\ref{eq:rho_nfw}
is fully specified by two parameters: a characteristic radius,
$r_s$, and a characteristic overdensity, $\delta_{\rm c}$. This simple
formula provides acceptable fits to DM halos independent of mass, 
cosmological paramters, fluctuation power spectra, and even holds
in several modified gravity models
\citep[e.g. NFW;][]{ColeLacey1996,Huss1999,Bode2001,Bullock2001,Maccio2007,Neto2007,Knollmann2008,Wang2009,Hellwing2013}.
The characteristic radius defines the scale at which the logarithmic slope
of the density profile has the isothermal value of -2,
i.e. $d\ln\rho/d\ln r|_{r_{-2}}=-2$. For that reason we will use $r_s$
and $r_{-2}$ interchangeably.

More recent simulation work confirmed the original findings of NFW,
but highlighted the need for a third parameter to accurately
capture the subtle halo-to-halo variation in profile shape
\citep[e.g.][]{Navarro2004,Hayashi2008,Navarro2010} as
well as its slight but systematic dependence on mass
\citep{Gao2008,Ludlow2013,Dutton2014}. Improved fits are obtained with 
profiles whose logarithmic slopes follow simple power-laws with
radius, $d\ln\rho/d\ln r\propto r^{-\alpha}$ 
\citep[e.g.][]{Merritt2005,Merritt2006}. One example is 
the Einasto profile \citep{Einasto1965}, which may be written as
\begin{equation}
  \ln\biggr(\frac{\rho_{\rm E}(r)}{\rho_{-2}}\biggl)=-\frac{2}{\alpha}\biggr[\biggr(\frac{r}{r_{-2}}\biggl)^\alpha-1\biggl].
  \label{eq:rho_einasto}
\end{equation}
An Einasto profile with $\alpha\approx 0.18$ closely resembles
the NFW over several decades in radius \citep[see, e.g.][]{Navarro2010}.

The scaling parameters of the NFW and Einasto profiles can be
expressed in alternative forms, such as halo virial\footnote{We
  define the virial mass of a halo as that enclosed by a sphere
  (centred on the potential minimum) of mean density equal to 200
  times the critical density, $\rho_{\rm crit}=3H^2/8\pi G$, where $H(z)$
  is the Hubble constant; the virial radius is therefore implicitly
  defined by $200\rho_{\rm crit}={\rm M}_{200}/(4/3)\pi r_{200}^3$.
  Note that {\em all} particles in the simulation are used in
  calculating ${\rm M}_{200}$ and not only those deemed
  gravitationally bound to a particular halo.} mass, ${\rm M}_{200}$, and
concentration, $c\equiv r_{200}/r_{-2}$, defined as the ratio of the
virial radius to that of the scale radius. At a given halo mass, the
concentration provides an alternative measure of the characteristic
density of a halo.

As discussed by NFW, ${\rm M}_{200}$ and $c$ do not take on arbitrary
values, but correlate in a way that reflects the mass-dependence of
halo formation times: those that assemble earlier have higher
characteristic densities, reflecting the larger background density at
that epoch. They used this finding to build a simple analytic model
based on the extended Press-Schechter theory \citep[EPS;][]{BCEK1991}
that reproduced the average mass
and cosmology dependence of halo concentrations
in their early simulations. 

Subsequent work by \citet{Bullock2001} corroborated the general trends
reported by NFW, but underscored a much stronger redshift dependence
of the concentration-mass relation than expected from their
model. These authors proposed a revised model that predicts 
concentrations which, at fixed
mass, scale linearly with expansion factor ($c\propto (1+z)^{-1}$)
and, at fixed $z$, fall off rapidly with increasing mass.
Later numerical work, however, found a much weaker mass and redshift
dependence than predicted by this model. Most notably, the
concentrations of very massive halos are found to be approximately
constant and to evolve little with redshift
\citep{Zhao2003a,Gao2008,Zhao2009}.

Empirical models that link halo concentrations to the shape of their
assembly histories fare better. The models of \citet{Wechsler2002} and
\citet{Zhao2003a}, for example, assume that the concentration is set
by the changing accretion rate of a halo, with the characteristic
density tracing the time when halos transition from an initial rapid
accretion phase to a subsequent phase of slower growth.  Very massive
systems are still in their initial rapid-growth phase at present, thus
explaining why they all have similar concentrations.

More recently, \citet[][hereafter L13]{Ludlow2013} used the Millennium
simulations (hereafter MS) to investigate the connection between CDM
halo assembly and structure. They pointed out that halo mass profiles
and main-progenitor mass accretion histories (MAHs) are self-similar
and have similar shapes. This becomes apparent when expressing mass
profiles in terms of average enclosed density,
${\rm M}(\langle\rho\rangle(r))$, rather than radius, and MAHs in terms of
main progenitor mass as a function of cosmic density rather than time
or redshift, i.e.  ${\rm M}(\rho_{\rm crit}(z))$. Both follow the NFW
profile. The two ${\rm M}(\rho)$ functions can thus be linked by a simple
scaling that allows the characteristic density of a halo (or its
concentration) to be inferred from the critical density of the
Universe at a characteristic time along its MAH.  \citet[][hereafter
L14]{Ludlow2014} showed how this result can be used to build a simple
analytic model for the mass-concentration-redshift relation that
accurately reproduced the trends obtained for CDM halos in a large
number of simulations, as well as the cosmology dependence of
$c({\rm M},z)$ reported in previous work.

Although the model works well for CDM \citep[see,
e.g.,][]{Correa2015c}, its applicability to models with truncated
power-spectra, such as those expected for ``warm'' dark matter (WDM),
is unclear. Interest in such models has been revitalized by recent
claims of detection of a $\sim$ 3.5 keV X-ray line, which is in
principle consistent with the decay of a {\rm warm} dark matter
particle \citep[e.g.][but see, Malyshev, Neronov \& Eckert 2014;
Anderson, Churazov \& Bregman 2015]{Bulbul2014,Boyarsky2014,Boyarsky2015}.
These results have motivated observations of dark matter dominated dwarf
galaxies of the Local Group \citep{Lovell2015} which, to date, have
not provided compelling evidence for dark matter decay \citep{Jeltema2015}.

The structure of WDM halos has been studied by \citet{Maccio2013} and,
more recently, by \citet{Bose2016}, who report that the NFW formula
accounts well for their mass profile shape 
\citep[see also, e.g.,][]{Knebe2002,VND2011,Polisensky2015,Gonzalez2016}. 
The resulting $c({\rm M})$
relation, however, is non-monotonic: concentrations reach a maximum at
a halo mass about 2 decades above the truncation scale but decline
gradually towards larger and smaller masses 
\citep[see also, e.g.,][]{Aurel2012,Maccio2013}. 
This implies that fairly
massive WDM halos can be as concentrated as low-mass ones although
the shape of their MAHs differ strongly, a feature that is difficult
to reconcile with the MAH-based scenario discussed above for CDM.

\begin{center}
 \begin{table*}
   \caption{Numerical aspects of our runs. $V_{\rm box}$ 
     is the simulation volume; N$_{\rm p}$ the total number of particles; $\epsilon$ the 
     Plummer-equivalent gravitational force softening; $m_{\rm p}$ the particle
     mass, and $m_{\rm WDM}$ the assumed mass of the thermal WDM particle
     and M$_{\rm{ hm}}$ the half-mode mass of its fluctuation power spectrum
     (left blank for cold dark matter models). For the case of the
     \textsc{coco} and Aquarius simulations, $V_{\rm box}$, N$_{\rm p}$,
     $\epsilon$ and $m_{\rm p}$ refer only to the high-resolution region.}
   \begin{tabular}{l c c c c c c c c c c}\hline \hline
     Simulation   & Model   &  $V_{\rm box}$      &  N$_{\rm p}$    &
     $\epsilon$     &  $m_{\rm p}$             & $m_{\rm WDM}$&  M$_{\rm{ hm}}$&\\
     &         &  [$h^{-3}\,{\rm Mpc}^3$]  &           &
     [$h^{-1}\,$kpc]& [$h^{-1}\,$M$_{\odot}$] & [keV]    &
     [$10^{10}\, h^{-1}\,$M$_{\odot}$] \\\hline
     \textsc{coco}& CDM/WDM & 2.2$\times 10^4$    &  2344$^3$ & 0.23           & 1.135$\times 10^5$      & 3.3 &0.025&\\\vspace{0.25cm}
     \textsc{color-1.5} & WDM     & 3.5$\times 10^5$    &  1620$^3$ & 1              & 6.20$\times 10^6$       & 1.5  &0.34&\\
     MS-II        & CDM     & 1$\times 10^6$      &  2160$^3$ & 1              & 6.89$\times 10^6$       & -&-&\\
     MS-I         & CDM     & 1.3$\times 10^8$    &  2160$^3$ & 5              & 8.61$\times 10^8$       & -&-&\\
     MS-XXL       & CDM     & 2.7$\times 10^{10}$ &  6720$^3$ & 10             & 6.17$\times 10^9$       & -&-&\\\vspace{0.25cm}
     Aquarius     & CDM/WDM &  $-$                &  $\sim 809^3$ &
     0.05       & 1.09$\times 10^4$       & 1.5, 1.6, 2.0, 2.3 & 0.34, 0.28, 0.13, 0.08&\\
     Cosmo-A      & CDM     & 5.2$\times 10^{7}$  &  1080$^3$ & 7.5            & 1.72$\times 10^9$       & -&-&\\
     Cosmo-B      & CDM     & 8.6$\times 10^{6}$  &  1080$^3$ & 4.1            & 4.78$\times 10^8$       & -&-&\\
     Cosmo-C      & CDM     & 1.1$\times 10^{7}$  &  1080$^3$ & 4.5            & 7.22$\times 10^8$       & -&-&\\
     Cosmo-D      & CDM     & 5.5$\times 10^{6}$  &  1080$^3$ & 3.5            & 4.84$\times 10^8$       & -&-&\\
   \end{tabular}
   \label{TabSimParam}
 \end{table*}
\end{center}

Indeed, there have been relatively few attempts
to model the $c({\rm M},z)$ relation for truncated power spectra. One
exception is the work of \citet{Eke2001}, who assumed that both the
normalization and {\em shape} of the power spectrum modulates
$c({\rm M},z)$. By including a term proportional to
$d\ln\sigma/d\ln {\rm M}$ in the definition of the collapse time they were
able to reproduce the concentration-mass relation in both CDM models
as well as several with truncated power spectra. This particular
parametrization, however, does not lend itself to simple
interpretation and predictions of their model were not borne out by
more recent simulations \citep[e.g.,][]{Gao2008,Diemer2015}.
\citet{Aurel2015} provide empirical relations
that can be used to map $c(M,z)$ relations obtained for CDM halos
to those expected for warm or mixed DM models.

It is clear that a full picture of DM halo structure must address
not only the mass and cosmological parameter dependence of halo
concentrations, but also the effect of the initial density fluctuation
spectrum. As more and more observations sensitive to the small-scale
clustering of DM become available, theoretical tools such as these
will be indispensable. This is the focus of the current paper.
Using a large suite of CDM and WDM cosmological simulations we
study the relationship between halo assembly and structure, paying
particular attention to the signature of a potential WDM particle.

Our paper is structured as follows. We begin with a description of our
simulations in Section~\ref{SecSims}; their analysis (including halo
finding, merger tree construction and density profile estimates) are
outlined in Section~\ref{SecCatalogs}. In Section~\ref{SecRes} we
present our main results. The mass-concentration-redshift relation
and its connection to halo assembly are presented in
Sections~\ref{sSec_cmz} and \ref{sSec_MAH}. We then use these results
to build a simple analytic model for $c({\rm M},z)$ in
Section~\ref{sSec_model}, which is compared to other available models
in Sections~\ref{sSec_model_comp_cc} and \ref{sSec_model_comp}. 
Finally, in Section~\ref{SecConc}, we provide a summary of our findings.
We elaborate on various aspect of our $c({\rm M},z)$ model in 
Appendix~\ref{AppComp} and \ref{AppModel} and provide an accurate 
fitting function for CDM halos in the Planck cosmology in Appendix~\ref{AppPlanck}.

\section{Numerical Simulations}
\label{SecSims}

Our analysis focuses on the growth histories and internal structure of
collisionless dark
matter halos identified in a suite of cosmological numerical
simulations. The majority of our results are based on the Copernicus
Complexio (\textsc{coco}) simulations \citep{Hellwing2016,Bose2016},
supplemented by the Millennium 
\citep{Springel2005a,Boylan-Kolchin2009,Angulo2012} and Aquarius simulations
\citep{Springel2008b,Lovell2014}, and an additional suite of $\Lambda$CDM 
runs which vary the parameters of the background cosmological
model\footnote{Various aspects of the post-processed simulation data
  may be made available by the first author upon request.}. The main 
aspects of these models are detailed in Tables~\ref{TabSimParam} and 
\ref{TabCosmoParam}. We provide here a brief description of the runs, but 
refer the reader to the original papers for a more thorough discussion. 

Note that in each of our WDM simulations we can safely neglect the 
intrinsic thermal velocities of the particles which, at $z=0$, are of
order a few tens of $m s^{-1}$. These particles will free-stream
only a few kiloparsecs over a Hubble time, which is comparable to our
interparticle separation \citep[see][]{Lovell2012}.

\subsection{The Copernicus Complexio simulations}
\label{SecCOCO}

The \textsc{coco} simulations track the evolution of dark matter in an
approximately spherical high-resolution volume of radius $\sim$
$18\Mpch$ embedded within a lower-resolution periodic box of
side-length $70.4\Mpch$. The high-resolution region contains
approximately 13 billion particles and was chosen in order to provide
a cosmologically representative sample of Milky Way-mass halos whilst
avoiding the unnecessary computational overhead of including
substantially more massive systems. To this end, the high-resolution
region was selected so that: 1) it includes no halos more massive than
$5\times 10^{13}\, h^{-1} {\rm M}_{\odot}$; 2) has no halos more massive
than $5\times 10^{14}\, h^{-1} {\rm M}_{\odot}$ within $\sim 5\Mpch$ of its
boundary, and 3) has a number density of $\sim 10^{12}\, h^{-1} {\rm M}_\odot$
halos that matches the universal halo mass function.

Linear perturbations were generated at $z=127$ using second-order
Lagrangian perturbation theory \citep{Jenkins2013} assuming a standard $\Lambda$CDM power
spectrum, as well as with a truncated power spectrum compatible with a
3.3 keV thermal WDM particle. We will hereafter refer to these runs as
\textsc{coco-cold} and \textsc{coco-warm}, respectively. Both simulations have identical 
phases and cosmological parameters, the latter adopting values consistent
with the WMAP 7-year data release \citep{Komatsu2011}: $\Omega_{\rm M}=0.272$,
$\Omega_{\Lambda}=0.728$, $\sigma_8=0.81$, $h=0.704$, and
$n_{\rm s}=0.967$. Here $\Omega_i$ is the present-day contribution to the
energy density from component $i$; $\sigma_8$ the linearly extrapolated rms
mass-fluctuation in spheres of 8\Mpch; $h$ is the current Hubble
expansion rate in units of $100{ \, \rm km \, s^{-1} \, Mpc^{-1}}$;
and $n_{\rm s}$ is the primordial spectral index of density
perturbations. With these choices of cosmological and numerical
parameters, the high-resolution particle-mass in the \textsc{coco} runs is
$m_{\rm p}= 1.135\times 10^5\, h^{-1} {\rm M}_{\odot}$.

We have also run a lower resolution version of the \textsc{coco-warm} simulation
assuming a lighter WDM particle of mass 1.5 keV. We will refer to this
run as \textsc{color-1.5} (\textsc{COco-LOw Resolution}, 1.5 keV). This run adopts the same 
set of WMAP-7 cosmological parameters, but samples the full $70.4\Mpch$ box with
1620$^3$ particles of equal mass, $m_{\rm p}=6.2\times 10^6\, h^{-1} {\rm M}_{\odot}$. 
We will use this run to assess the effect of changing the thermal cut-off in
the dark matter power spectrum on the internal structure of DM halos.

\subsection{The Millennium and Aquarius simulations}
\label{SecMill}

Because of the relatively small volume of the \textsc{coco} simulations, we
will extend the dynamic range of our analysis using 
the Millennium simulation suite. Each of these runs 
adopts cosmological parameters which were chosen to match
the WMAP-1 $\Lambda$CDM values -- $\Omega_{\rm M}=0.25$;
$\Omega_{\Lambda}=0.75$; $\sigma_8=0.9$; $h=0.73$; $n_{\rm s}=1$ -- but
differ in both total particle number and in box size. The
Millennium \citep{Springel2005a} and Millennium-II \citep{Boylan-Kolchin2009}
simulations evolve the dark matter density field using ${\rm N_p}=2160^3$
particles in periodic boxes with side-lengths equal to 500 and
100\Mpch, respectively. The Millennium-XXL simulation
\citep{Angulo2012} adopts both a larger
particle number, ${\rm N_p}=6720^3$, and box-size, 
${\rm L_{box}}=3\,h^{-1}{\rm Gpc}$, than either MS or MS-II. 

We also use the latest suite of CDM and WDM simulations from the
Aquarius Project \citep{Lovell2014}.  Like \textsc{coco}, the Aquarius
simulations assumed a WMAP-7-normalized power spectrum, but focused
computational resources on a single Milky Way-mass
dark matter halo and its surroundings. Each run has the same
high-resolution particle mass,
$m_{\rm p}=1.09\times 10^4\, h^{-1} {\rm M}_{\odot}$ (equivalent to level-2
in the original nomenclature of the Aquarius Project), and identical
phases of the initial Gaussian random field, but adopt transfer
functions appropriate for CDM and thermal WDM models of mass
$m_{\rm WDM}=2.3, 2.0, 1.6$ and 1.5 keV.

\subsection{Additional runs}
\label{SecCosmo}

We have also carried out four additional flat $\Lambda$CDM simulations which
vary the parameters of the background cosmological model. Each run
uses 1080$^3$ equal-mass particles, assumes $h=0.73$ and $n_{\rm s}=1$, but
varies the matter density parameter, $\Omega_{\rm M}$, and rms fluctuation
amplitude, $\sigma_8$. The cosmological parameters of these runs, which we
have labelled Cosmo-A, B, C and D, are provided in Table~\ref{TabCosmoParam}.

\begin{center}
 \begin{table}
   \caption{Parameters for the cosmological models studied in 
     this paper. Each run is a flat $\Lambda$CDM or $\Lambda$WDM cosmology. $\Omega_{\rm bar}$, 
     $\Omega_{\rm M}$ and $\Omega_\Lambda$ are the present-day energy
     densities in baryons, total matter and cosmological constant, respectively; $h$
     is the Hubble parameter, expressed in units of $100{ \, \rm km \,
       s^{-1} \, Mpc^{-1}}$; $\sigma_8$ is the rms linear density fluctuation
     in 8\Mpch~ spheres; and $n_{\rm s}$ the power-law index of the primordial
     density fluctuation spectrum.}
   \begin{tabular}{l c c c c c c c}\hline \hline
     Model   &$\Omega_{\rm bar}$& $\Omega_{\rm M}$ & $\Omega_\Lambda$ & $h$   & $\sigma_8$ & $n_{\rm s}$ \\\hline
     Planck  &   0.0484         &  0.308     & 0.692            & 0.678 & 0.815      & 0.968 \\
     WMAP-7  &   0.0446         &  0.272     & 0.728            & 0.704 & 0.81       & 0.967 \\
     WMAP-1  &   0.045          &  0.25      & 0.75             & 0.73  & 0.9        & 1.0   \\
     Cosmo-A &   0.045          &  0.15      & 0.85             & 0.73  & 1.0        & 1.0   \\
     Cosmo-B &   0.045          &  0.25      & 0.75             & 0.73  & 0.6        & 1.0   \\
     Cosmo-C &   0.045          &  0.29      & 0.71             & 0.73  & 0.81       & 1.0   \\
     Cosmo-D &   0.045          &  0.40      & 0.60             & 0.73  & 0.7        & 1.0   \\
   \end{tabular}
   \label{TabCosmoParam}
 \end{table}
\end{center}

\begin{figure*}
  \includegraphics[width=0.85\textwidth]{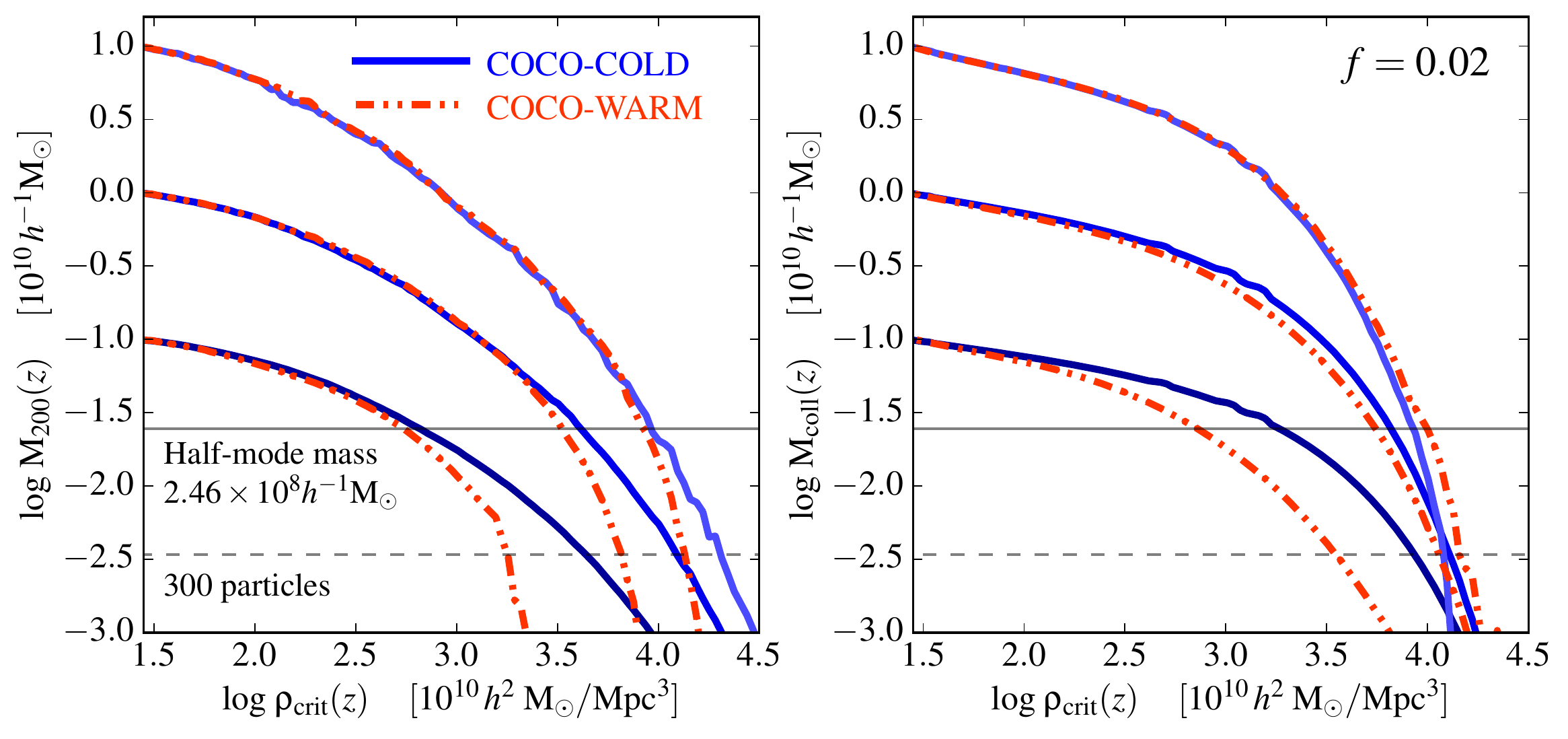}
  \caption{Median assembly histories of halos identified at $z_0=0$,
    grouped in three narrow mass bins centered at ${\rm M}_0=10^9$,
    $10^{10}$, and $10^{11}\, h^{-1}\, {\rm M}_\odot$. Blue (solid)
    lines correspond to halos identified in \textsc{coco-cold} and
    (dot-dashed) orange lines to \textsc{coco-warm}. The left panel
    shows the mass accretion histories (MAH), defined as the evolution
    of the virial mass, ${\rm M}_{200}(z)$, of the main
    progenitor. The right-hand panel shows the {\em total} mass,
    ${\rm M}_{\rm coll}(z)$, in collapsed progenitors with mass exceeding
    $2\%$ of ${\rm M}_0$. The (solid) grey horizontal line marks the
    ``half-mode'' mass, at which the WDM transfer function is
    suppressed by a factor of two relative to CDM.}
  \label{fig:mz_mrho_unscaled}
\end{figure*}

\section{Halo Inventory and Analysis Techniques}
\label{SecCatalogs}

\subsection{Halo identification}
\label{SecSubfind}

All simulation outputs were processed with a friends-of-friends (FoF)
group finder \citep{Davis1985} using a linking-length of $b=0.2$ times the
Lagrangian inter-particle separation; at each snapshot, groups with fewer than 20
particles were discarded. The substructure finder \textsc{subfind}
\citep{Springel2001b} was then run
on the remaining FoF groups in order to identify their self-bound
subhalos. \textsc{subfind} decomposes each FoF group into a dominant (or central)
subhalo and a contingent of less-massive subhalos
that trace the self-bound relics of past accretion events. For simplicity,
we will refer to the assemblage of each central halo and its full subhalo 
population as a ``main halo''. 

For each main halo, \textsc{subfind} records a virial mass, ${\rm M}_{200}$, and associated
radius, $r_{200}$. For our analysis we will retain only main halos that
exist as distinct objects at $z_0=0,1,2$ or 3 and, additionally, contain at least 
${\rm N}_{200}=5000$ particles within their virial radius.

\subsection{Assembly histories}
\label{SecMAH}

The halo catalogue is used to construct merger trees for each 
main halo following the procedure described in \citet{Jiang2014}.
This method tracks particles within each 
subhalo across simulation outputs in order to determine their descendants.
Subhalos and their descendants are then split into unique branches, with
new branches growing when a subhalo first appears in the simulation and
continuing until it has fully merged with a more massive system. The
merger tree of a particular halo is then constructed by packaging
the \textsc{subfind} merger trees of each of its surviving subhalos.

Using these merger trees we construct mass accretion histories for each 
halo, defined as the evolution of the virial mass, ${\rm M}_{200}(z)$, of its 
{\em main progenitor} (hereafter MAH). The left panel of Fig.~\ref{fig:mz_mrho_unscaled}
shows, for halos identified at $z_0=0$, the median MAHs computed in three separate
mass bins. Solid (blue) curves correspond to \textsc{coco-cold}; dot-dashed (orange) curves
to \textsc{coco-warm}. Note that the MAHs of CDM and WDM halos differ
strongly below the characteristic mass scale imposed by the free-streaming
of the WDM particle, shown here as a horizontal grey line at the
``half-mode'' mass\footnote{The half-mode mass, ${\rm M}_{\rm hm}$, indicates the scale at
  which the WDM transfer function is reduced by half relative to a
  CDM model with the same cosmological parameters. In our
  \textsc{coco-warm} model this corresponds to ${\rm M}_{\rm hm}=2.46\times
  10^8 \, h^{-1}\, {\rm M}_{\odot}$, and in \textsc{color-1.5} to ${\rm M}_{\rm
    hm}=3.4\times 10^9 \, h^{-1}\, {\rm M}_{\odot}$. Half-mode masses
  for our remaining WDM runs are provided in Table~\ref{TabSimParam}.}.

Although the MAH provides a useful proxy for the assembly history of a
halo, it neglects the full spectrum of progenitors that contribute to
its growth, motivating alternative measures. One possibility is to
tally the mass of {\em all} progenitors that have collapsed by
redshift $z$ and that are above some fraction $f$ of the halo's final mass,
${\rm M}_0$. This quantity, referred to as the ``collapsed mass
history'' (CMH) and denoted ${\rm M}_{\rm coll}(z)$,
has a simple interpretation and is easily extracted from simulated or
theoretical DM merger trees.

The right-hand panel of Fig.~\ref{fig:mz_mrho_unscaled} shows
the CMH for $z_0=0$ halos in the same mass bins as
those used in the left-hand panel. Solid (blue) curves again
indicate \textsc{coco-cold} halos and dot-dashed (orange)
\textsc{coco-warm}; all assume $f=0.02$. Unlike ${\rm M}_{200}(z)$,
these curves do not show a characteristic suppression of growth below
the WDM free-streaming scale. This may be readily understood using the
EPS theory to compute analytically the collapsed mass fraction 
\citep[e.g.][]{Lacey1993}:
\begin{equation}
{\rm M}_{\rm coll}(z) = {\rm M}_0 \times \erfc\biggr(\frac{\delta_{\rm
    sc}(z)-\delta_{\rm sc}(z_0)}{\sqrt{2\,
  (\sigma^2(f\, {\rm M}_0)-\sigma^2({\rm M}_0))}}\biggl),
\label{eqMcoll}
\end{equation}
where $\delta_{\rm sc}(z)\approx 1.686/D(z)$ is the $z=0$ density threshold for
the collapse of a spherical top-hat perturbation, $D(z)$ is the
linear growth factor, and $\sigma({\rm M})$ the rms density fluctuation in
spheres enclosing mass ${\rm M}$. Note that the redshift dependence of this
function enters only in the numerator; its {\em shape} therefore
depends only on the background expansion history, independent of
$\sigma({\rm M})$ or $f$. As we will see in
Section~\ref{sSec_cmz}, this has important consequences for models of
the $c({\rm M},z)$ relation that relate the characteristic densities
of DM halos to their MAH-based formation times.

\begin{figure*}
  \includegraphics[width=0.7\textwidth]{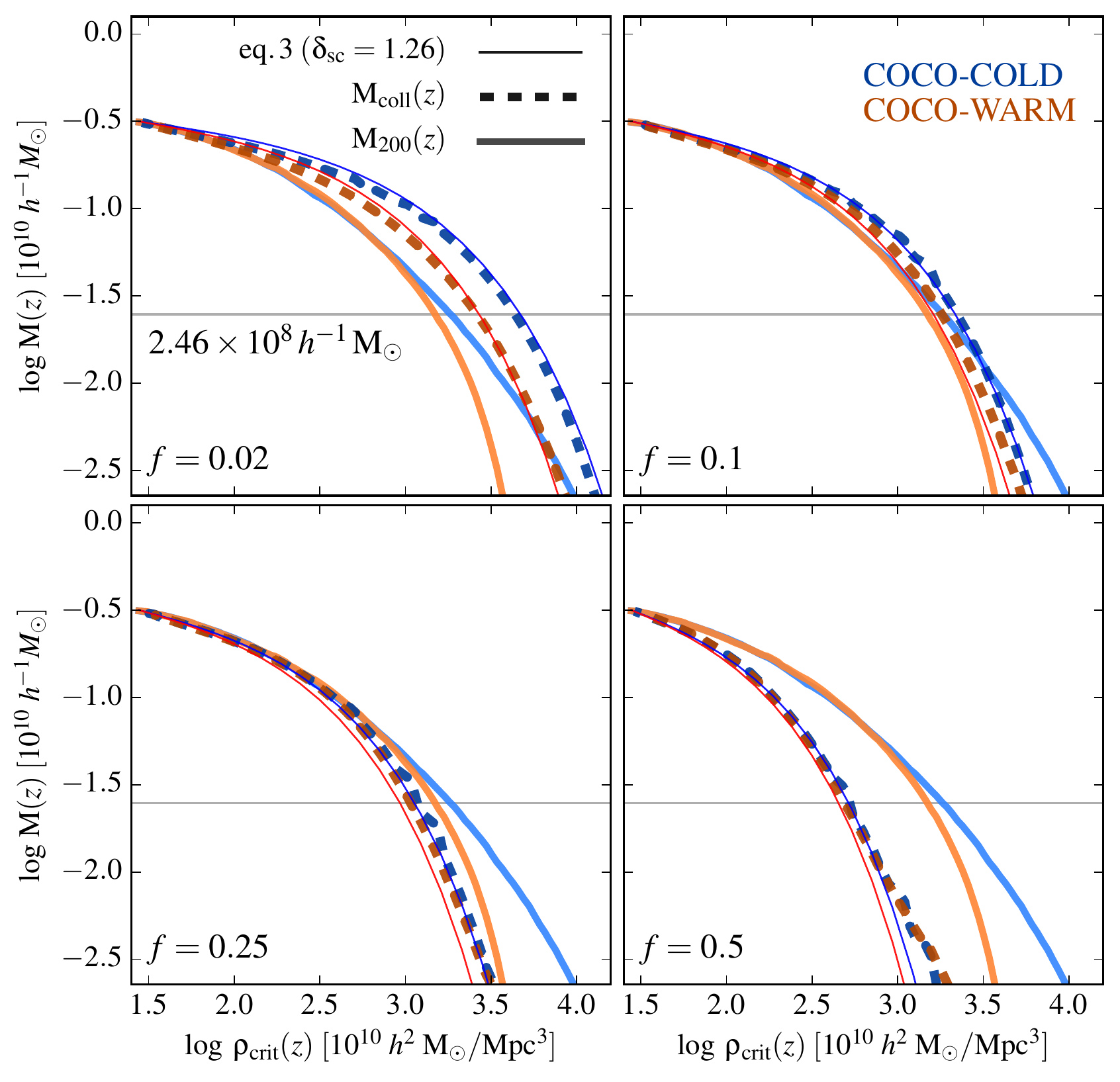}
  \caption{Median mass assembly histories of CDM and WDM halos of mass
    ${\rm M_0}= 3.2\times 10^9\, h^{-1}\, {\rm M}_\odot$. Thick dashed
    lines in each panel show ${\rm M_{coll}}(z)$, defined as the total
    mass in progenitors more massive than $f\times M_0$. Different
    panels show results obtained for different values of $f$. Also
    shown in each panel are the mass accretion histories, ${\rm
      M_{200}}(z)$, which track the evolution of the virial mass of
    the main progenitor. Below the free-streaming ``half-mode'' mass,
    shown here as a horizontal grey line, the evolution of ${\rm
      M_{200}}(z)$ differs substantially between the CDM and WDM
    models. The shape of ${\rm M_{coll}}(z)$, on the other hand, is the
    same for both, independent of the value of $f$. The thin solid
    lines show the shapes of these curves anticipated from EPS theory
    (see eq.~\ref{eqMcoll}), which have a fixed shape and describe all
    curves reasonably well.}
  \label{fig:mz_mcoll}
\end{figure*}

The thick dashed lines in Fig.~\ref{fig:mz_mcoll} show the CMH of a
$3.2\times 10^9\, h^{-1}\, {\rm M}_\odot$ halo for several values of
the parameter $f$, and compares them to the median MAH (solid lines,
repeated in each panel to aid the comparison).  Thin lines show
eq.~\ref{eqMcoll}, adopting $\delta_{\rm sc}=1.26$ for the collapse
threshold\footnote{We choose a value for $\delta_{\rm sc}$ lower than
  the canonical value of $1.686$ in order to account for inaccuracies
  of the spherical collapse model \citep[e.g.][]{Sheth2001,LBP2014}. This choice does not
  alter the shape of $\rm{M_{coll}}(z)$.}. Note that this expression
describes the shape of $\rm{ M_{coll}}(z)$ remarkably well for both
\textsc{coco-cold} and \textsc{coco-warm}, independent of the value of
$f$ adopted. Unlike the main-progenitor MAH,
the CMH provides a {\em universal} description of the halo
assembly process, where the choice of $f$ implicitly defines the halo
collapse time: lower values of $f$ imply earlier formation redshifts.

\subsection{Mass profiles and concentration estimates}
\label{SecMrho}

For each halo identified at $z_0=0,1,2$ and 3 we have constructed
spherically-averaged density profiles, $\rho(r)$, in $32$
equally-spaced steps in $\log r$ spanning $-2.5\leq \log r/r_{200}
\leq r_{200}$. Within each radial bin we also compute the total
enclosed mass, ${\rm M}(r)$, and mean inner density profiles,
$\langle\rho\rangle(r)={\rm M}(r)/(4/3)\pi r^3$. To ensure that our halo
mass profiles are well resolved, we restrict our analysis to those
with ${\rm N}_{200}\geq 5000$ particles within their virial radius,
$r_{200}$.

We construct the $c({\rm M},z)$ relation by fitting {\em median} mass
profiles after averaging over logarithmic mass bins of width
$\Delta\log {\rm M}=0.1$. This smooths out any features unique to individual systems 
and dampens the influence of outliers allowing for a robust estimate of the 
average mass and redshift-dependence of halo concentrations. 

In practice, best-fit concentrations are determined by adjusting the
three parameters of eq.~\ref{eq:rho_einasto} in order to minimize a 
figure-of-merit, defined
\begin{equation}
  \psi^2 = \frac{1}{{\rm N}_{\rm bin}} \sum_{i=1}^{\rm Nbin} [\ln \rho_i - \ln \rho_E(\rho_{-2};r_{-2};\alpha)]^2,
  \label{eq:FoM}
\end{equation}
over the radial range $r_{\rm conv}<r<0.8\, r_{200}$. Here $r_{\rm conv}$ 
is the \citet{Power2003} convergence radius corresponding to the median
mass profile; the outer limit of $0.8\, r_{200}$ ensures that our fits
exclude radii that may not be fully relaxed \citep[see, e.g.][for
a full discussion]{Ludlow2010}. Because we fit
{\em median} profiles, statistical errors in the density estimates for
individual radial bins are extremely small and may be neglected.

Einasto profiles can also be expressed in terms of the enclosed density:
\begin{equation}
  \langle\rho_{\rm E}\rangle(r)=\frac{{\rm M}(<r)}{(4\pi/3)r^3}=\frac{200}{x^3}\frac{\Gamma(3/\alpha;2/\alpha
    \,(x c)^\alpha)}{\Gamma(3/\alpha;2/\alpha \, c^\alpha)}\rho_{\rm{crit}},
  \label{eq:Mrho}
\end{equation}
where $x=r/r_{200}$ and $\Gamma(a;y)$ is the incomplete
$\Gamma$-function. Halo mass profiles,
expressed in terms of their enclosed density, ${\rm M}(\langle\rho\rangle)$,
can be fit with eq.~\ref{eq:Mrho} to provide an alternative measure of
concentration. To do so, we first normalize the mass and density 
profiles by their present-day values, ${\rm M_0=M}_{200}(z_0)$ and 
$\rho_0=\rho_{\rm crit}(z_0)$, and determine the remaining 
parameters, $c$ and $\alpha$, by minimizing the rms deviation between
$M(\langle\rho\rangle)$ and eq.~\ref{eq:Mrho}. After some experimentation,
we found concentrations estimated this way to be less susceptible to moderate changes in
the adopted radial fit range when applied to {\em individual} halos, whilst
leaving the median trends unchanged. For that reason, we adopt 
this method whenever individual fits are required.

\begin{figure*}
  \includegraphics[width=0.9\textwidth]{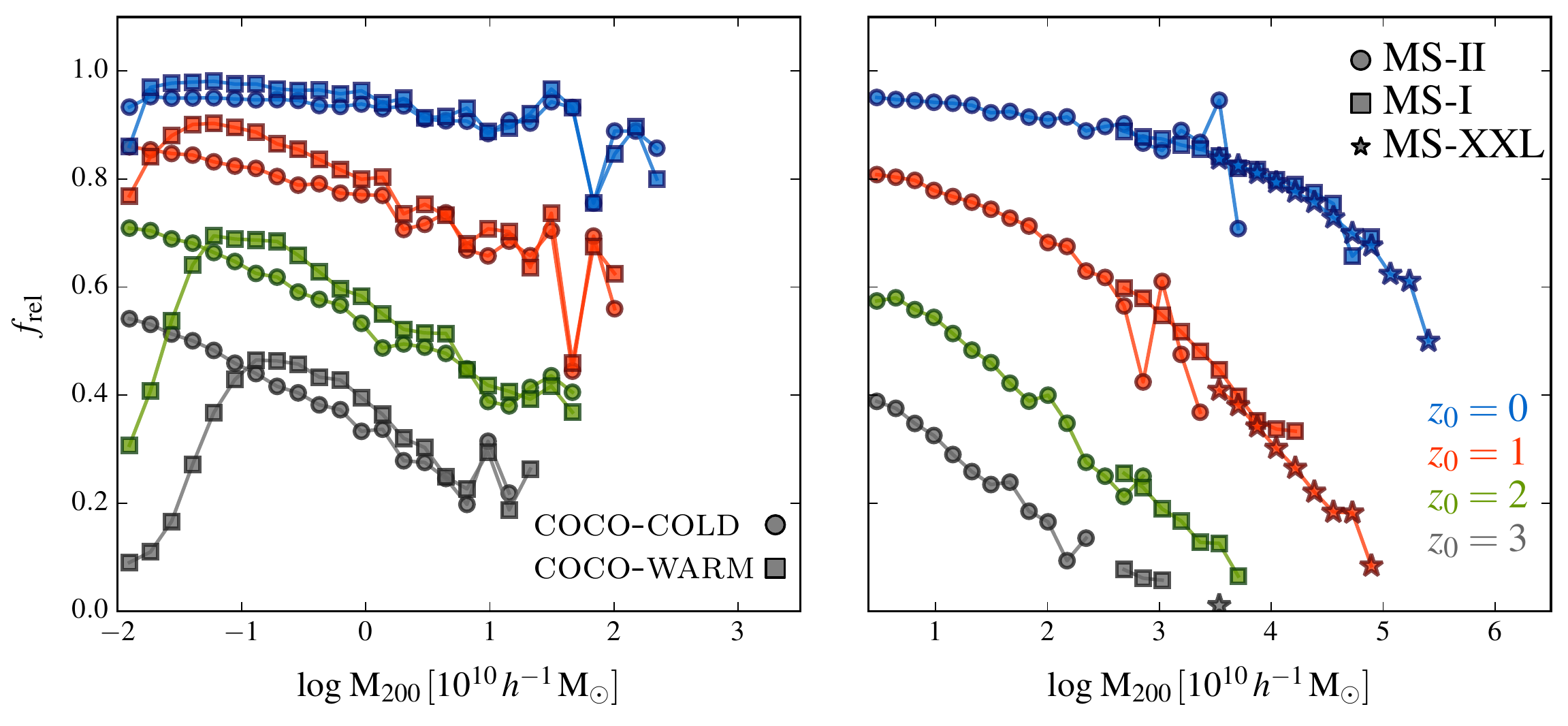}
  \caption{Fraction of relaxed halos as a function of mass and redshift for the
    \textsc{coco} (left) and Millennium simulations (right). Relaxed halos are defined
    as those for which $t_{50}\simgt 1.25\times t_{\rm cross}$ and $d_{\rm off} <0.1$. Here,
    $t_{50}$ is the look-back half-mass formation time of the halo, $t_{\rm cross}$ is its
    crossing time and $d_{\rm off}$ the centre-of-mass offset parameter. Different 
    coloured lines correspond to different redshifts and point styles to different simulations,
    as indicated in the legend. For all CDM models, the relaxed fraction decreases with increasing
    mass, as expected for hierarchical cosmologies in which halo collapse times
    decrease monotonically with mass. Note, however, that, for WDM models, collapse 
    times are non-monotonic, resulting in a turnover in the relaxed fracton toward
    low masses, where formation times begin to decrease.}
  \label{fig:relx}
\end{figure*}

\subsection{Relaxed versus unrelaxed halos}
\label{SecRelaxed}

Dark matter halos form hierarchically through a combination of smooth
accretion, minor mergers and occasional major mergers with systems of
comparable mass. These events can drive large but transient departures
from quasi-equilibrium states during which the structural properties
of DM halos are rapidly evolving and ill defined. As a result, the
majority of studies aimed at calibrating the $c({\rm M},z)$ relation
have taken steps to identify and excise halos believed to be far from
equilibrium, thereby defining samples of ``relaxed'' halos with smooth
mass profiles whose structural features can be meaningfully described
with a few parameters. It is important to note, however, that relaxed
halos form a highly biased sub-sample of the full halo population, and
their prevalence depends in non-trivial ways on both halo mass and on
redshift \citep[see, e.g.,][]{Thomas2001,Maccio2007,Neto2007,Power2012,Ludlow2012,Angel2016,Klypin2016}.

\begin{figure*}
  \includegraphics[width=0.75\textwidth]{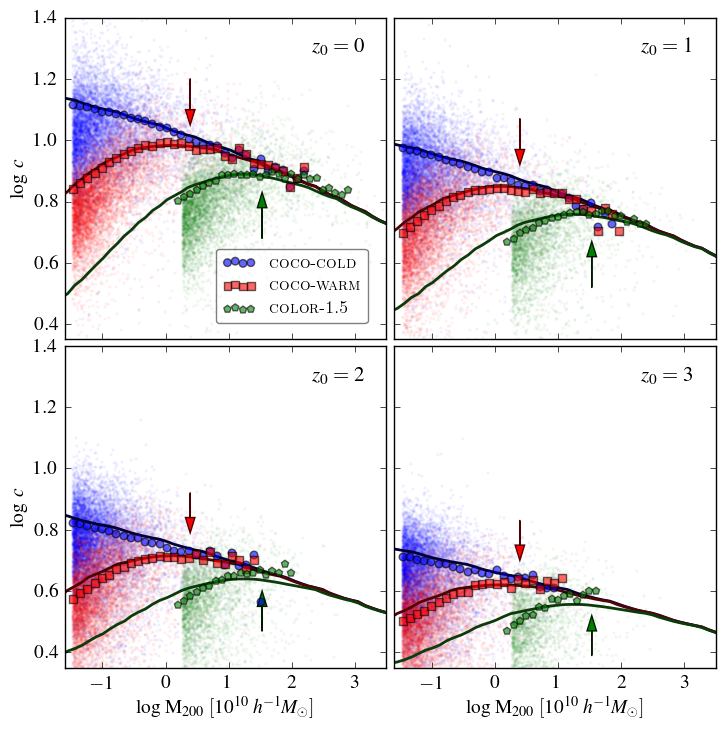}
  \caption{Mass-concentration-redshift relation for \textsc{coco-cold}
    (blue), \textsc{coco-warm} (red), and \textsc{color-1.5} (green).
    Fits to individual halos are shown as coloured dots; heavy points
    correspond to the best-fit relation derived from the {\em stacked}
    mass profiles of halos in equally-spaced logarithmic mass bins of
    width $0.1$ dex. Only bins containing at least 25 halos are shown.
    Note that in WDM model the concentration-mass relation is
    non-monotonic, and peaks at a mass scale that is a multiple of the
    free-streaming scale associated with the WDM particle. The
    arrows indicate a mass scale corresponding to one hundred
    times the half-mode mass, where the
    WDM power spectrum is suppressed by 50 per cent relative to CDM.
    The solid lines correspond to the predictions of the analytic
    model described in Section \ref{sSec_model} (see text for
    details).}
  \label{fig:cmz}
\end{figure*}

Identifying relaxed halos is not without ambiguity, and
a number of diagnostics have been proposed in the literature.
Some are sensitive to geometric halo properties, such as the
centre-of-mass offset parameter, defined as $d_{\rm off}=|\mathbf{r}_p-\mathbf{r}_{\rm
  CM}|/r_{200}$ \citep{Thomas2001,Maccio2007,Neto2007}, or the
mass-fraction in substructure \citep[e.g.][]{Neto2007,Ludlow2012};
others, such as the virial ratio, $\eta=2\, K/|U|$
\citep[e.g.][]{ColeLacey1996,Bett2007,KP2008}, or spin parameter, 
$\lambda$ \citep[e.g.][]{Klypin2016}, gauge the internal dynamical state 
of the halo. Some authors reject halos whose spherically average
density profiles are poorly described by their chosen fitting 
formulae \citep[e.g.][]{Maccio2007,Maccio2008,Dutton2014}.

\citet{Neto2007} suggested a combination of three criteria that may be
used to curtail halos whose mass profiles are most likely to deviate
from smooth spherical averages. These include: (i) the centre-of-mass
offset, $d_{\rm off}<0.07$, (ii) the substructure mass fraction,
$f_{\rm sub}=M_{\rm sub}(<r_{200})/{\rm M}_{200}<0.1$, and (iii) the
virial ratio, $\eta<1.35$. During a merger each of these quantities
fluctuate in predictable ways: $d_{\rm off}$, for example, traces the
centre-of-mass of a merging system about its densest core, providing
an estimate of the accuracy with which the halo centre can be defined;
$f_{\rm sub}$ monitors the mass contribution from undigested mergers,
while $\eta$ is sensitive to fluctuations in the gravitational
potential as orbital energy is dissipated into binding energy.  As
discussed by \citet{Ludlow2012} and also \citet{Poole2016},
merger-driven oscillations in these quantities are out of sync, making
it unlikely that a halo will simultaneously fail all three at any
point during a merger.

Are these criteria sufficient to ensure removal of all unrelaxed
halos? Arguably not. Because of its resolution dependence, only halos
with well-resolved substructure populations are sensitive to
$f_{\rm sub}$. In simulations with uniform mass resolution (such as
those used in this work) the least resolved halos are also the most
abundant; $f_{\rm sub}$ is therefore a useful equilibrium statistic
for only the most massive, best-resolved systems. Furthermore, since
DM halos are not truly isolated, the virial ratio should be corrected
for external forces and surface pressure terms \citep[see, e.g.,][for
a discussion]{pooleg2006,KP2008,Klypin2016} and only then can it be
used to meaningfully assess departures from equilibrium.

Because of these uncertainties we here adopt a simpler approach and
use the {\em dynamical age} of a system as the primary diagnostic for
equilibrium.  We assume that any halo whose main progenitor has more
than doubled in mass in under a crossing time cannot have had time to
relax to an equilibrium configuration. More specifically, we require
$t_{50}\simgt 1.25\times t_{\rm cross}$ as a minimal but necessary
condition for equilibrium. Here
$t_{\rm cross}\equiv 2\times r_{200}/{\rm V}_{200}$ is the
characteristic crossing time of a halo and $t_{50}$ is the lookback
half-mass formation time of its main progenitor. This single
criterion, however, will only flag halos undergoing very rapid
accretion or equal-mass mergers. For that reason, we additionally
impose the familiar criterion $d_{\rm off}<0.1$ to cull the remainder
of the population. Our primary motivation for
choosing these criteria is that they do away with uncertainties
surrounding the importance of boundary terms in the virial ratio, and the
resolution-dependence of $f_{\rm sub}$. We will see in Appendix~\ref{AppComp} that
imposing these criteria on MS halos results in a $c({\rm M},z)$
relation that decreases monotonically with mass over the redshift
range probed by our simulations, removing the ``upturn'' in the
concentration of high-mass halos reported by \citet{Klypin2011} 
(see \citet{Ludlow2012} for further details). 

It is worth noting, however, that different
definitions of what consitutes a relaxed halo population lead to 
conflicting claims regarding the origin of the upturn, and whether or not
it is a true property of the underlying structure of equilibrium DM halos 
\citep[see, e.g.,][for discussions]{Ludlow2012,Correa2015c,Klypin2016}.
A full assessment will likely require a detailed study of the
perils and virtues of a variety of different equilibrium benchmarks,
which we defer to future work.

The mass and redshift dependence of the relaxed
halo fraction, $f_{\rm rel}$ (defined above) is shown in Fig~\ref{fig:relx}. The left panel shows results 
for the \textsc{coco} simulations and the right for the Millennium simulations, 
which extend to much higher masses. In CDM models halo collapse times decrease 
monotonically with increasing mass, which is reflected in the decreasing 
abundance of relaxed halos amongs massive systems. For WDM models, however,
collapse times are not monotonic: there is a {\em maximum} formation time for
halos at any given redshift, resulting in a non-monotic relation between
halo mass and the prevalence of relaxed systems.

\begin{figure*}
  \includegraphics[width=0.8\textwidth]{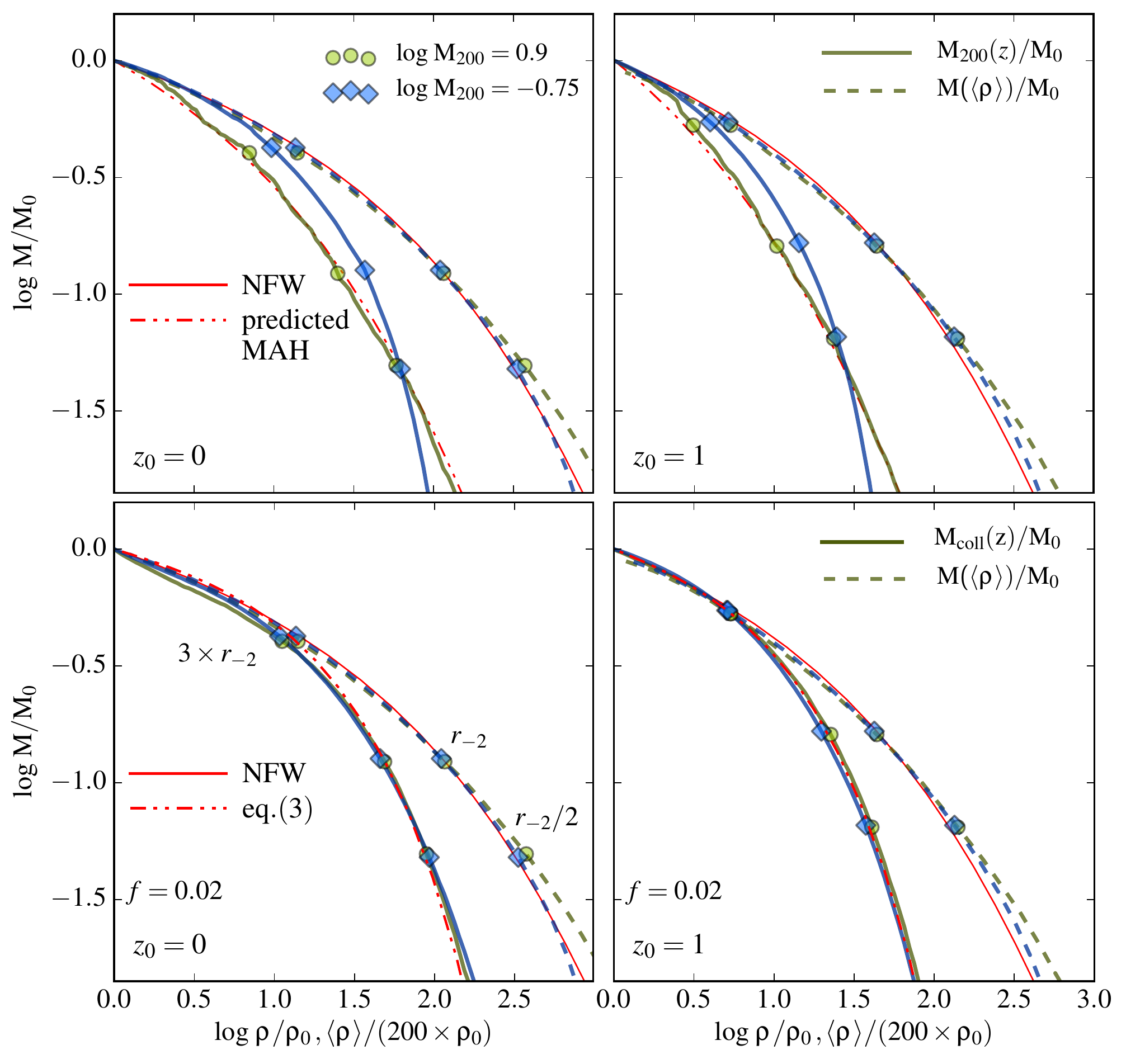}
  \caption{Enclosed density profiles, ${\rm M}(\langle\rho\rangle)$ (dashed
    lines) and assembly histories (solid lines) for halos in
    \textsc{coco-warm} at redshift $z=0$ (left panels) and $z=1$
    (right panels). Different coloured lines correspond to two
    separate mass bins: $\log {\rm M}_{200}/[10^{10} h^{-1} {\rm
      M}_\odot]=0.9$ (green curves) and $\log {\rm M}_{200}/[10^{10}
    h^{-1} {\rm M}_\odot]=-0.75$ (blue curves), chosen to have similar
    concentration but to fall on opposite sides of the ``peak''
    concentration. Note that, at each redshift, the median mass
    profiles of both halos (dashed lines) are remarkably similar.  The
    solid lines in the top panels show the main-progenitor MAH, ${\rm
      M}_{200}(z)$; those in the bottom panels show the total
    collapsed mass, $M_{\rm coll}(f,z)$, in progenitors more massive
    than $f \times M_0$, for $f=0.02$ (see text for details).  The
    outsized points along each curve mark the mass enclosed within
    $3\times r_{-2}$, $r_{-2}$ and $0.5\times r_{-2}$, as
    indicated. The solid red curves show NFW profiles with 
    concentration parameters equal to the median values, $c=9.5$ (left)
    and $c=6.4$ (right); the red
    dot-dashed curves in the upper panels show the NFW-MAH predicted
    using the procedure described in L14. The top panel shows that
    very different main-progenitor MAHs can lead to halos of the same
    concentration. The bottom panel shows that these same halos have
    very similar ``collapsed mass histories'', ${\rm M}_{\rm coll}(z)$.  The
    dot-dashed red curves in the lower panels indicate that ${\rm M}_{\rm
      coll}(z)$ is in excellent agreement with predictions based on
    EPS theory (eq.~\ref{eqMcoll}).}
  \label{fig:coco_mah}
\end{figure*}

\section{Results}
\label{SecRes}

\subsection{The $c({\rm M},z)$ relation in CDM and WDM}
\label{sSec_cmz}

The mass-concentration-redshift relations for equilibrium halos in the 
\textsc{coco} and \textsc{color-1.5} simulations are shown in
Fig.~\ref{fig:cmz}. Dots show the best-fit concentrations 
obtained for individual halos, colour-coded to distinguish different runs.
Symbols trace the median $c({\rm M},z)$ relations of the same sets of halos,
obtained by fitting the median mass profiles after averaging over 
logarithmic mass bins of width $\Delta\log {\rm M}=0.1$ (only bins 
containing at least 25 halos are plotted in this figure).

These results confirm and extend previous work on the structure of WDM
halos.  Unlike CDM, where concentrations increase monotonically with
decreasing mass, in WDM models the $c({\rm M},z)$ relation has a
characteristic shape: it first increases with decreasing mass, but
reaches a well-defined maximum before decreasing again towards lower
mass. Note that in the 3.3 and 1.5 keV models studied here the mass
scale at which the peak concentration is reached is roughly
independent of redshift. Note also that differences between WDM and
CDM are already evident at mass scales substantially larger than the
free-streaming scale.  For example, the ``peak'' in the median
concentration of WDM halos occurs approximately {\em two orders of
  magnitude above the half-mode mass}, suggesting that differences in
the very early stages of halo growth leave a lasting imprint on the
final halo. We highlight this point using coloured arrows, which 
correspond to {\em one hundred times} the half-mode mass of each WDM run. 
This provides an important clue for models that aim to fully describe
the $c({\rm M},z)$ relations from the power spectrum alone. The solid
curves in Fig.~\ref{fig:cmz} show the predictions of such a model,
which we describe in more detail in Section~\ref{sSec_model}.

\begin{figure*}
  \includegraphics[width=0.8\textwidth]{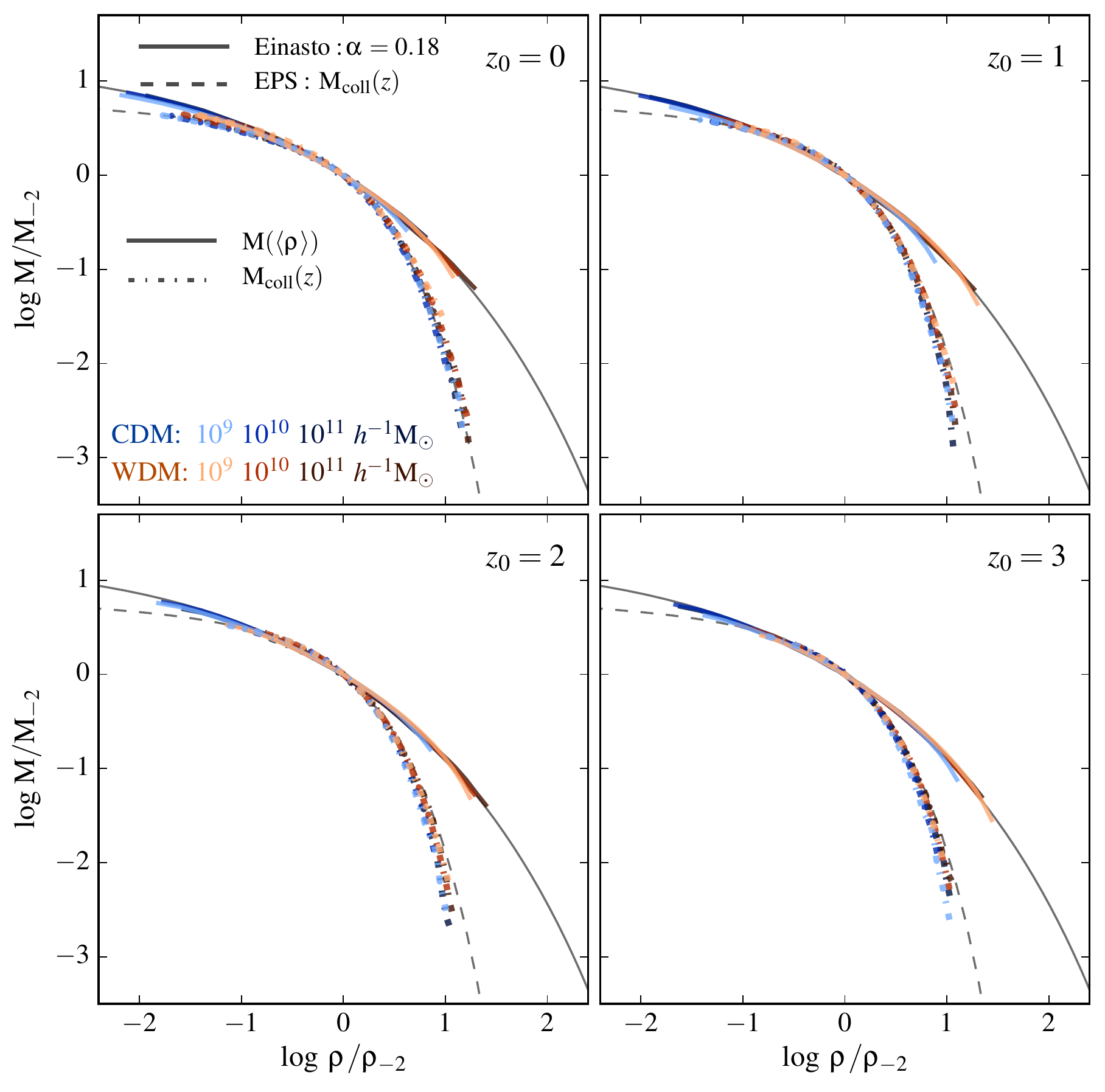}
  \caption{Median mass profiles (solid lines) and collapsed mass histories,
    ${\rm M}_{\rm coll}(z)$ (dot-dashed lines), for halos in three
    separate mass bins (${\rm M}_0=10^9$, $10^{10}$, and $10^{11}\,
    h^{-1} {\rm M}_\odot$) after rescaling to the characteristic values of
    mass, ${\rm M}_{-2}={\rm M}(<r_{-2})$, and density,
    $\rho_{-2}$. Separate panels show halos in the same mass bins, but
    identified at four different redshifts: $z_0=0$, $1$, $2$ and
    $3$. The thin grey line in each panel shows an Einasto profile
    with $\alpha=0.18$; dashed lines show eq.~\ref{eqMcoll}; i.e., the ${\rm
      M}_{\rm coll}(z)$ shape predicted by extended Press-Schechter
    theory.}
  \label{fig:mz_mrho_rescaled}
\end{figure*}

\subsection{Mass profiles and assembly histories}
\label{sSec_MAH}

As discussed by L13, the MAHs and mass profiles of cold dark matter
halos have, on average, the same NFW shape. This implies that halo
concentrations can be obtained by simply rescaling their mass accretion
history by a fixed amount: the characteristic density of
${\rm M}(\langle\rho\rangle)$ is simply proportional to that of 
${\rm M}(\rho_{\rm crit}(z))$. One consequence of this result is that
two halos of
similar MAH must have the same characteristic density and vice versa,
independent of their mass or identification redshift. This greatly
simplifies the task of predicting concentrations from assembly
histories when applied to CDM \citep[see also][]{Correa2015c}.  In WDM
models, however, the suppression of gravitational collapse below the
free-streaming scale breaks the scale-invariance of the assembly
process: it imprints a preferred scale on the MAHs, readily seen in
the left-most panel of Fig.~\ref{fig:mz_mrho_unscaled}. This implies
that the mass profiles of WDM halos {\em cannot be obtained by simply
  rescaling the MAHs of their main progenitors}.

We illustrate this in the upper panels of Fig.~\ref{fig:coco_mah},
where we show the median main-progenitor MAHs, ${\rm M}_{200}(z)$
(solid lines), and the enclosed density profiles, ${\rm
  M}(\langle\rho\rangle)$ (dashed lines), of \textsc{coco-warm} halos
for two different halo masses and at two different redshifts. The halo
masses ($\log {\rm M}_{200}/[10^{10} h^{-1} {\rm M}_\odot]=0.9$ and
$\log {\rm M}_{200}/[10^{10} h^{-1} {\rm M}_\odot]=-0.75$) were
selected so that their median concentration is roughly the same
($c\approx 9.5$ at $z_0=0$, and $c\approx 6.4$ at $z_0=1$), but fall
on opposite sides on the ``peak'' in the $c({\rm M})$ relations.  To aid the
comparison, masses have been normalized to the current mass, ${\rm
  M}_0={\rm M}(z_0)$, critical densities to $\rho_{\rm crit}(z_0)$, and
enclosed densities to $200\, \rho_{\rm crit}(z_0)$.

The dashed curves indicate that these halos not only have the same
$c$, but also similar mass profiles across the entire resolved radial
range. The outsized symbols highlight the enclosed mass and mean
density within $r=3\times r_{-2}$, $r_{-2}$ and $r_{-2}/2$, which are
roughly equivalent for both masses. For comparison, the solid red line
shows an NFW profile with the same concentration.

Despite the similarity of the halo mass profiles,
it is clear from Fig.~\ref{fig:coco_mah} that the {\em shapes} of the
MAHs of the two halos (solid lines) are substantially different. 
The dot-dashed curve shows the MAH obtained by rescaling the NFW mass
profile, as described in L13 for CDM. This model describes quite
well the MAH of massive WDM halos, but fails dramatically at low
mass, where the MAH shape differs substantially from NFW. The MAH of
such halos cannot be used then to infer the concentration of their mass
profile in the same way as for CDM halos. 

Alternative descriptions of halo growth that preserve scale invariance
may improve matters. One possibility, mentioned in
Section~\ref{SecMAH}, is to use the mass, ${\rm M}_{\rm coll}(f,z)$, in {\em
  all} progenitors (above a certain threshold $f\times {\rm M}_0$) rather
than just that of the main progenitor. The lower panels of
Fig.~\ref{fig:coco_mah} show, for the same two halo masses, the growth
of the total mass in progenitors more massive than 2 per cent of the
halo's final mass. The curves are now virtually indistinguishable,
suggesting that the collapsed mass in progenitors {\em other} that the
main one plays an important role in establishing a halo mass profile.
Note also that the shape of ${\rm M}_{\rm coll}(z)$ is accurately described
by eq.~\ref{eqMcoll}, shown in the lower panels using a dot-dashed
(red) line after rescaling to match each of the halo formation
times. This suggests that it may be possible to use the collapsed mass
history, ${\rm M}_{\rm coll}(f,z)$, to predict halo concentrations. 

Fig.~\ref{fig:mz_mrho_rescaled} shows that this is indeed the case,
for both CDM and WDM halos, regardless of mass or identification
redshift. Here we compare the median ${\rm M}_{\rm coll}(z)$
(dot-dashed lines), constructed using $f=0.02$ for halos of mass ${\rm
  M}_0=10^9$, $10^{10}$ and $10^{11} \, h^{-1} {\rm M}_\odot$ after
rescaling each to the characteristic values of mass, ${\rm
  M}_{-2}={\rm M}(<r_{-2})$, and density, $\rho_{-2}$. 

Different panels show results for different identification redshifts,
with blue and orange curves distinguishing halos in the
\textsc{coco-cold} and \textsc{coco-warm} runs, respectively.  As
anticipated, each curve has a similar shape, independent of ${\rm
  M}_0$, $z_0$ or the shape of the DM power spectrum. Indeed, as
alluded to above, the similarity of these curves is actually {\em
  expected} from EPS theory.  The dashed grey line in each panel of
Fig.~\ref{fig:mz_mrho_rescaled}, for example, shows eq.~\ref{eqMcoll},
which describes the shapes of these curves remarkably well.

Fig.~\ref{fig:mz_mrho_rescaled} also shows the median enclosed
density profiles, ${\rm M}(\langle\rho\rangle)$, for each set of
halos, again after normalizing to the characteristic values of ${\rm
  M}_{-2}$ and $\langle\rho_{-2}\rangle$. The solid grey line shows an
Einasto profile with $\alpha=0.18$, which
  provides an accurate approximation to the median mass profiles of
  these halos over the mass and redshift range probed by our
  simulations\footnote{It is worth noting that the best-fit values of 
    $\alpha$ obtained for halos in the \textsc{coco} simulations are consistent with previously
    published results \citep[e.g.][]{Gao2008,Dutton2014,Klypin2016},
    but remain relatively constant due to the limited mass range probed by these runs.}.

Because both ${\rm M}_{\rm coll}(\rho_{\rm crit}(z))$ and
${\rm M}(\langle\rho\rangle)$ have self-similar (albeit distinct) shapes, a
single scaling relation between characteristic density and formation
time is sufficient to anchor the two and to construct an analytic
model for $c({\rm M},z)$, provided that, for some fixed value of $f$, the
characteristic density of the CMH is
simply proportional to that of the mass profile.

\begin{figure}
  \includegraphics[width=0.4\textwidth]{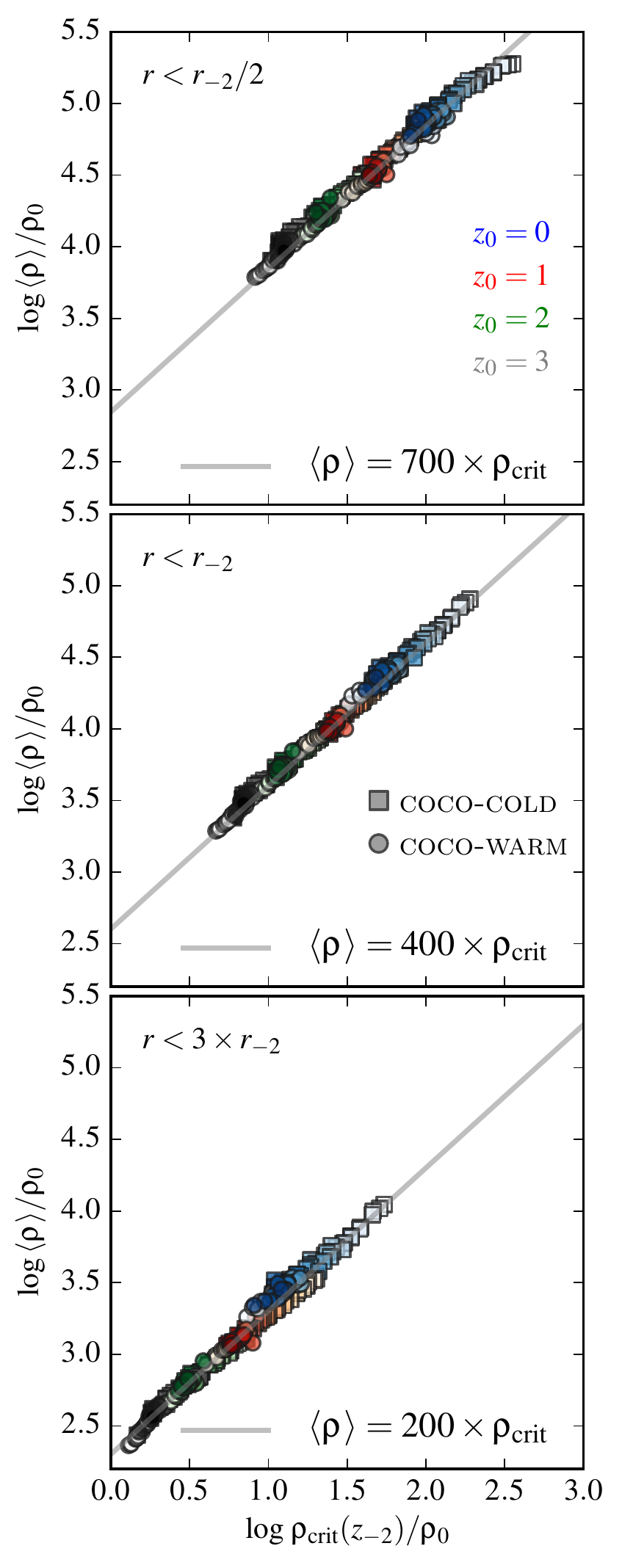}
  \caption{Mean enclosed density within $r_{-2}$ (middle), 
    $r_{-2}/2$ (top) and $3\times r_{-2}$ (bottom), versus the
    critical density at the time when the mass enclosed by each of those radii
    was first contained in progenitors more massive than $0.02\times {\rm M}_0$.
    In each case, enclosed and critical densities have been scaled to 
    the current value, $\rho_0=\rho_{\rm crit}(z_0)$.}
  \label{fig:rhorho}
\end{figure}

We show this in Fig.~\ref{fig:rhorho}, where we plot the
mean enclosed density within the halo scale radius,
$\langle\rho_{-2}\rangle=M_{-2}/(4/3)\pi r_{-2}^3$, versus $\rho_{\rm
  crit}(z_{-2})$, the critical density at the redshift $z_{-2}$ when
the enclosed mass was first assembled into progenitors more massive
than 2 per cent of ${\rm M}_0$ (i.e., $f=0.02$).  Note that the linear scaling,
$\langle\rho_{-2}\rangle=C\times \rho_{\rm crit}(z_{-2})$, between
these two densities is independent of both mass and identification
redshift. More importantly, however, the zero-point of this relation
is independent of the DM particle model: both \textsc{coco-cold} and
\textsc{coco-warm} have $C\approx 400$.  Note also that similar
scalings can be found for characteristic densities measured within
different fractions of $r_{-2}$. This can be seen in the upper and
lower panels of Fig.~\ref{fig:rhorho} which show the results of
repeating this calculation within $r_{-2}/2$ and $3\times r_{-2}$,
respectively.

\begin{figure}
  \includegraphics[width=0.5\textwidth]{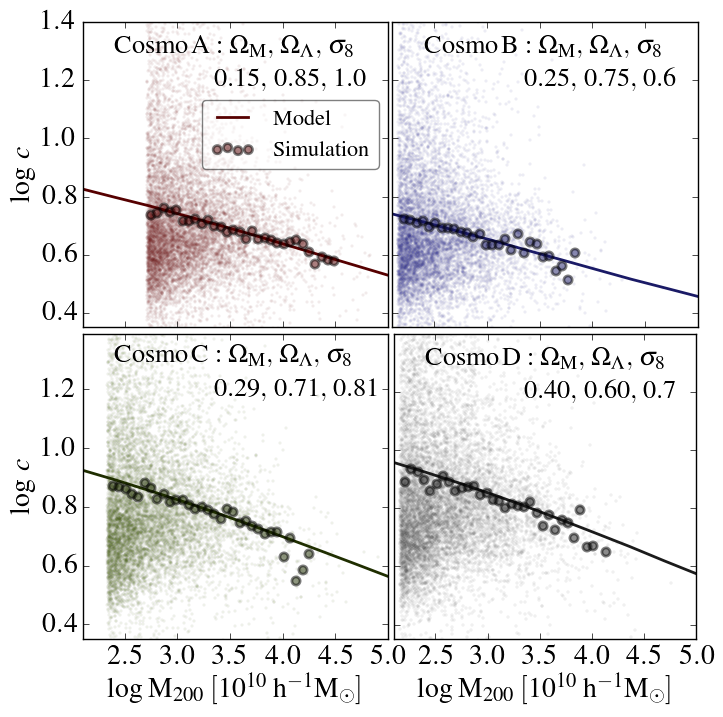}
  \caption{Concentration-mass relation for equilibrium halos at 
    $z_0=0$ in four runs which vary the background cosmological 
    parameters $\Omega_{\rm M}$ and $\sigma_8$. Coloured dots correspond 
    to individual halos; filled symbols to the median $c({\rm M})$ relation 
    computed in uniformly spaced logarithmic mass bins. Solid curves show 
    the predictions of the model described in Section
    \ref{sSec_model} with parameters $f=0.02$ and $C=650$. }
  \label{fig:cmz_cosmo}
\end{figure}

\subsection{An analytic model for the $c({\rm M},z)$ relation}
\label{sSec_model}

The results of the previous section suggest that the $c({\rm M},z)$
relation can be inferred from collapsed mass histories, ${\rm M}_{\rm
  coll}(z)$, provided those can be obtained from either simulations or
theoretical models. Encouragingly, the most recent generation of
algorithms accurately reproduce not only the evolution of the main
progenitor halo
\citep[e.g.][]{vandenBosch2002,Jiang2014b,Correa2015a,Correa2015b},
but the entire hierarchy of progenitors for both CDM
\citep[e.g.][]{Parkinson2008} and WDM \citep[e.g.][]{Benson2013}
fluctuation power-spectra.

In order to describe the {\em average} relation between concentration,
mass and redshift, we use analytic arguments based on EPS theory to
construct a simple model. Following the procedure laid out by NFW, the
model assumes that a halo's characteristic density reflects the
critical density of the Universe at a suitably-defined collapse
redshift. NFW adopted $\delta_c$ as the characteristic density
(eq.~\ref{eq:rho_nfw}), and defined the collapse redshift as the time
when half the {\it virial} mass of the halo was first contained in
progenitors more massive than some fraction $f$ of the final {\it
  virial} mass.  

A simple modification to this procedure yields much better results
which are applicable to both CDM and WDM initial power spectra. These
modifications identify the characteristic density with the mean inner
density within the scale radius, $\langle\rho_{-2}\rangle$, rather
than $\delta_c$, and the characteristic halo mass with ${\rm M}_{-2}$ rather
than the virial mass.

The revised model assumes that $\langle\rho_{-2}\rangle$ is directly 
proportional to the critical density of the Universe at the collapse 
redshift, $z_{-2}$, given by
\begin{equation}
  \frac{\langle\rho_{-2}\rangle}{\rho_0} \equiv C\times\frac{\rho_{\rm
      crit}(z_{-2})}{\rho_0}= C\times \biggr[\frac{H(z_{-2})}{H(z_0)}\biggl]^2.
  \label{eq:rho_scaling}
\end{equation}
With this definition, the collapse redshift denotes the redshift at
which the characteristic mass, ${\rm M}_{-2}$, was first contained in 
progenitors more massive than a fraction $f$ of the final halo mass. 
According to eq.~\ref{eqMcoll}, the collapse redshift can therefore be 
obtained from
\begin{equation}
  \frac{{\rm M}_{-2}}{{\rm M}_0}\equiv \erfc\biggr(
  \frac{\delta_{\mathrm sc}(z_{-2})-\delta_{\mathrm
      sc}(z_0)}{\sqrt{2(\sigma^2(f\times {\rm M}_0)-\sigma^2({\rm M}_0))}} \biggl).
  \label{eq:PS}
\end{equation}
Note that the left-hand sides of eqs~\ref{eq:rho_scaling} and \ref{eq:PS} depend
only on the halo mass profile; once $f$ and $C$ have been specified,
and the shape of the mass profile assumed, these equations can 
be solved simultaneously for the concentration as a function of mass, 
${\rm M}_0$, redshift, $z_0$, and its dependence on power spectrum through
$\sigma({\rm M})$\footnote{Throughout the paper we calculate $\sigma({\rm M})$
  from the linear power spectrum using a real-space spherical top-hat
  filter. Although other possibilities exist, our tests show that the best
  results are obtained with this choice.}. 

\begin{figure}
  \includegraphics[width=0.5\textwidth]{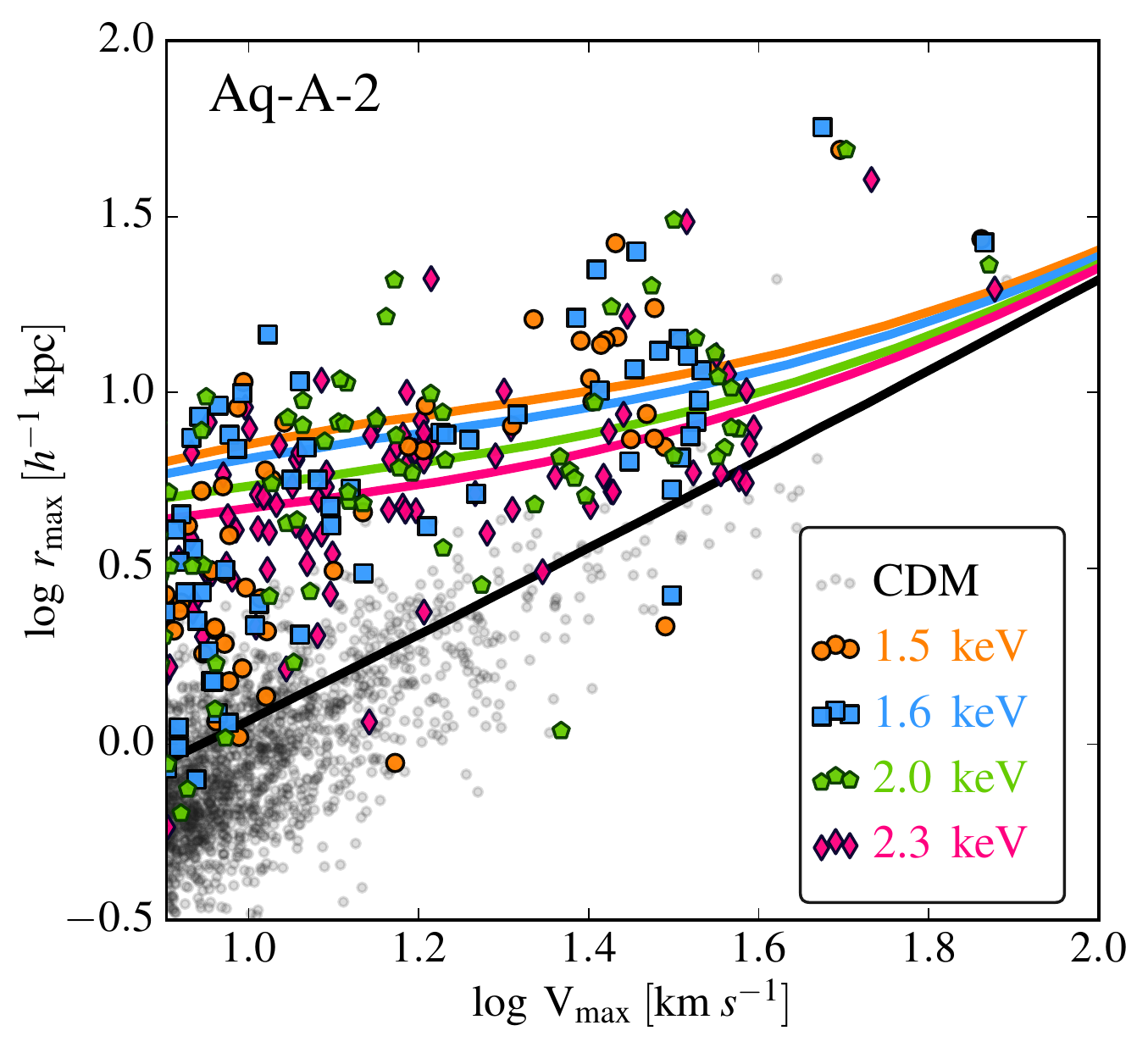}
  \caption{The relation between $r_{\rm max}$ and $V_{\rm max}$ for 
    isolated field halos (defined as those that lie farther than 
    2$\times \, r_{200}$ from the largest halo in the simulation) 
    in the level-2 Aquarius simulations of \citet{Lovell2014}. Black 
    dots show individual measurements for the CDM cosmogony, and 
    coloured symbols indicate runs with $m_{\rm WDM}=1.5$, 1.6, 2.0 
    and 2.3 keV. Solid lines of corresponding colour show the predicted
    relations using the model described in Section~\ref{sSec_model}. As in
    previous plots, the free parameters of the model are chosen to be
    $C=650$ and $f=0.02$.}
  \label{fig:vmax_rmax}
\end{figure}

The solid curves in Fig.~\ref{fig:cmz} show the resulting $c({\rm M},z)$ 
relations for $f=0.02$ and using a normalization
constant\footnote{Note that this zero point is different to $C=400$
  which is obtained from the simulations (see the middle panel of 
  Fig.~\ref{fig:rhorho}). This is due to inaccuracies of the spherical 
  collapse model, which is implicitly assumed in the calculation of 
  eq.~\ref{eq:PS}. Merger trees generated using Monte Carlo methods 
  {\em tailored} to reproduce simulations results
  \citep[e.g.][]{Parkinson2008} should adopt $C=400$ and $f=0.02$.} 
of $C=650$, assuming that $\rho(r)$ follows an Einasto\footnote{
    We show in Appendix~\ref{AppModel} that the predictions of the
    model depend only weakly on the assumed shape of the DM density profile,
    provided it resembles those of simulated halos. For simplicity, we
    have adopted a single Einasto profile with shape parameter
    $\alpha=0.18$ throughout the paper.} profile with $\alpha=0.18$.
Note that this model accurately reproduces the median 
trends for CDM, and both WDM cosmologies considered here without any 
additional tuning of the parameters.

Having calibrated the model parameters using these
runs, we can now make {\em predictions} for $c({\rm M},z)$ that
can be tested against simulations of different WDM and cosmological models.
One such test is shown in Fig.~\ref{fig:cmz_cosmo}, where we plot the
$c({\rm M})$ relations for equilibrium halos in four separate flat
$\Lambda$CDM cosmological models.  Each has ${\rm H}_0=73\, {\rm km}\,
{\rm s}^{-1}{\rm Mpc}^{-1}$ and $n_{\rm s}=1$, but varies the matter density
parameter, $\Omega_{\rm M}$, and rms fluctuation amplitude,
$\sigma_8$, as indicated in the legends. As in Fig.~\ref{fig:cmz},
dots show the best-fit concentrations for individual halos; filled
symbols show the median relations.  The solid lines show the
predictions of eqs.~\ref{eq:rho_scaling} and \ref{eq:PS} for $C=650$
and $f=0.02$, assuming an Einasto mass profile with $\alpha=0.18$. 
Note that these are the {\em same} parameters used to fit the
$c({\rm M},z)$ relations in Fig.~\ref{fig:cmz}, which were obtained
from simulations of a WMAP-7 cosmological model. 

A further test of the model involves its extrapolation to much lower
halo masses than can be resolved in \textsc{coco-cold} and \textsc{coco-warm}. We show
this in Fig.~\ref{fig:vmax_rmax} where we compare the $V_{\rm
  max}$-$r_{\rm max}$ relation (equivalent to the mass-concentration
relation) for field halos in the high-resolution region of the
Aquarius simulations of \citet{Lovell2014}. These runs follow the
evolution of Milky Way-mass dark matter halos and their immediate
surroundings in CDM and a series of WDM models with thermal particle
masses of $\approx \,$1.5, 1.6, 2.0 and 2.3 keV.  We plot here
isolated halos, defined as main halos that are farther than two virial
radii from the largest halo in the simulation. This selection ensures
that systems previously accreted and expelled from the most massive
halo in the simulation are excluded from the analysis
\citep[see][]{Ludlow2009}. Grey dots show the ${\rm V}_{\rm max}-r_{\rm
  max}$ relation obtained from the CDM run; coloured symbols show the
results for the various WDM models, as indicated in the legend.

The predictions of the analytic model described above are shown in
Fig.~\ref{fig:vmax_rmax} for all DM models considered. As in previous
plots, we compute each curve for $C=650$ and $f=0.02$; line and symbol
colours were chosen to match the corresponding simulations. These
curves accurately describe the systematic shift in concentration
brought about by changes to the properties of the DM particle. The
good agreement between model and simulation implies that our model
provides a useful theoretical tool for studies aiming to constrain the
nature of dark matter based on its imprint on the internal structure
of low-mass halos.

\begin{figure*}
  \includegraphics[width=0.8\textwidth]{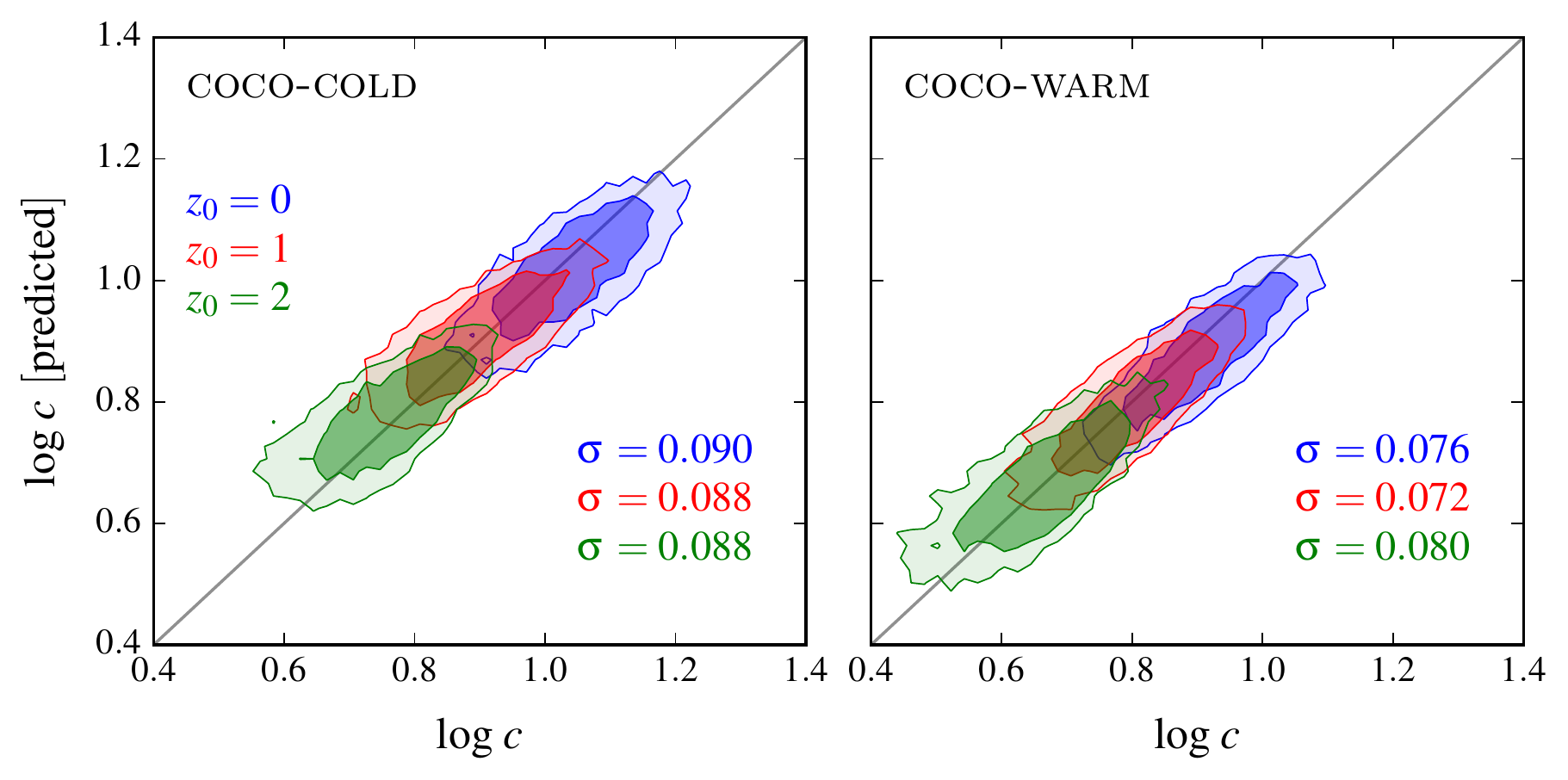}
  \caption{Comparison of the predicted and measured concentrations for
    equilibrium halos with ${\rm N}_{200}\geq 10^4$ particles in the 
    \textsc{coco-cold} (left) and \textsc{coco-warm} (right)
    simulations. Our predictions assume that each halo's
    characteristic density, $\langle\rho_{-2}\rangle$, is directly 
    proportional to the critical background density when the CMH 
    (for $f=0.02$) first crosses ${\rm M}_{-2}$. Concentration is 
    then calculated by assuming that each halo's mass profile follows
    an $\alpha=0.18$ Einasto profile.}
  \label{fig:cc_individ}
\end{figure*}

\subsection{Concentrations of individual halos}
\label{sSec_model_comp_cc}

Although eqs.~\ref{eq:rho_scaling} and \ref{eq:PS} provide a robust
account of the median $c({\rm M},z)$ relation in each of our simulations,
our methodology can also be applied to {\em individual} halos. To do so, we
select equilibrium systems with ${\rm N}_{200}>10^4$ particles from the
\textsc{coco-cold} and \textsc{coco-warm} simulations 
and use their merger trees to construct ${\rm M}_{\rm coll}(z)$. Assuming
that each halo's mass profile is reasonably well described by an
($\alpha=0.18$) Einasto profile, its concentration can be readily
obtained from eq.~\ref{eq:rho_scaling}, with $C=400$ in this case.

Fig.~\ref{fig:cc_individ} plots
the predicted versus the measured concentrations. 
The left-hand panel shows results for \textsc{coco-cold}, the
right-hand panel for \textsc{coco-warm}. The coloured contours enclose 75 and 50
per cent of the data points, and are shown for $z_0=0$, 1 and 2. This
figure makes clear that our procedure faithfully reproduces the
concentrations of relaxed halos. The rms scatter about the one-to-one
line, for example, is typically less than $\sim$0.09, indicating that the
error on the predicted value of $c$ is less than $\sim$23 per cent.

\begin{figure*}
  \includegraphics[width=0.75\textwidth]{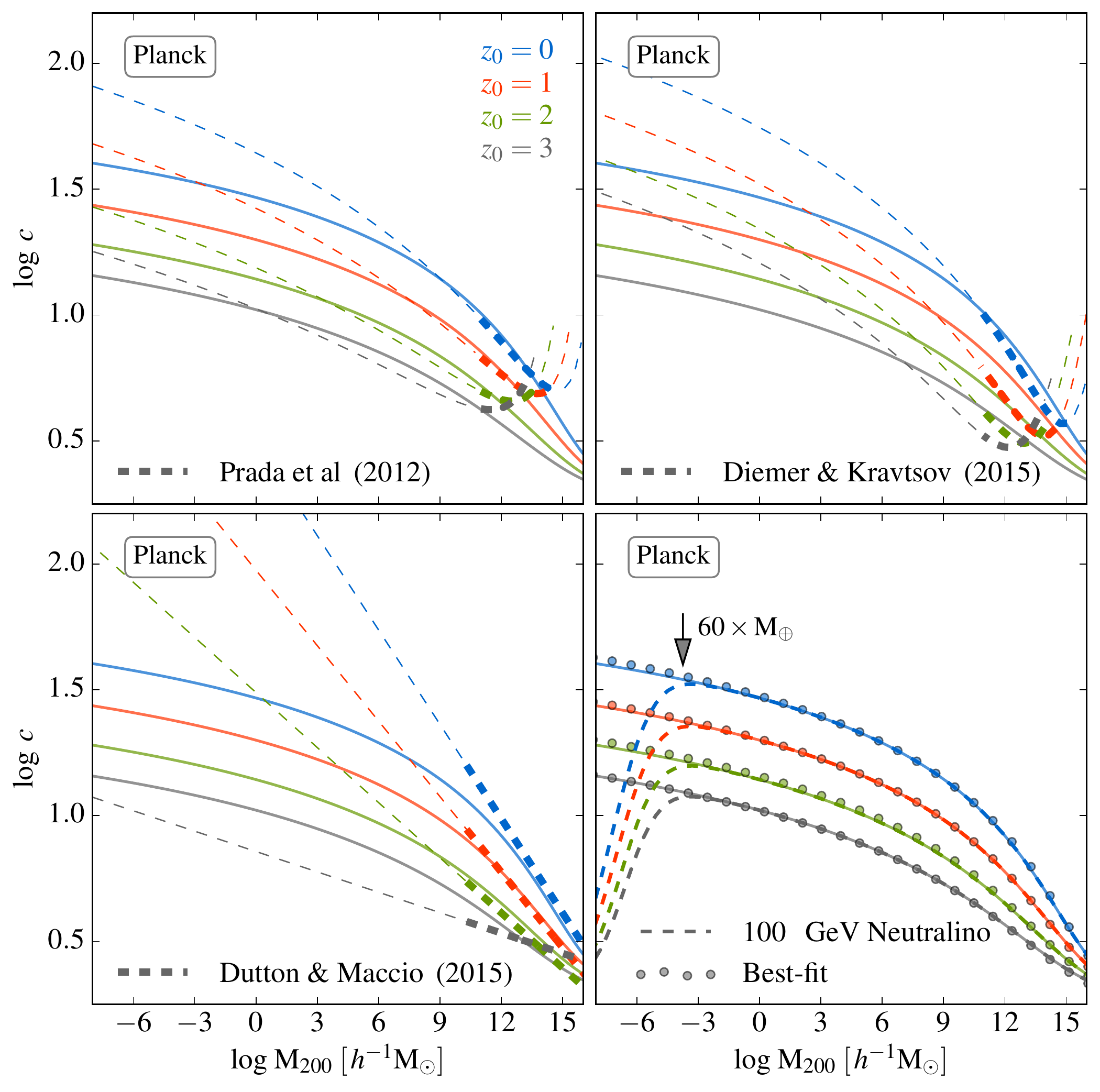}
  \caption{Comparison of the mass-concentration-redshift relation 
    in the Planck cosmology to those of: 1) \citet{Prada2012} (upper-left);
    2) \citet{Diemer2015} (upper-right), and 3) \citet{Dutton2014}
    (lower-left). Solid coloured lines (repeated in each panel) show
    the predictions of the model described in Section~\ref{sSec_model};
    other models are shown using dashed lines of similar colour,
    with thick segments highlighting the mass
    range over which they were calibrated. In the 
    lower-right panel, dashed lines show the predictions for a 100 GeV
    neutralino CDM model, and the filled symbols show our best-fitting
    function to the pure CDM case (see Appendix~\ref{AppPlanck} for details).
    The downward arrow marks sixty Earth masses, below which the concentrations
    of neutralino halos are highly suppressed.}
  \label{fig:cmz_fits}
\end{figure*}

The fraction of scatter in $c({\rm M})$ that is due to different halo
collapse times can be estimated by comparing the variance in the
measured and predicted values of concentration for halos in a given
mass bin. For CDM halos at $z_0=0$, we find that variation in $z_{-2}$
accounts for $\sim 40$ per cent of the scatter for ${\rm M_{200}} \sim
10^9 h^{-1} {\rm M}_\odot$, and $\sim 50$ per cent at $\sim 10^9
h^{-1} {\rm M}_\odot$. For WDM
halos of the same mass and redshift, the reductions are $\sim 80$
and $\sim 50$ per cent, respectively. Thus, the scatter in collapse
time does not fully account for the scatter in concentration. Future
studies should assess the role of environment or initial conditions in 
establishing the mass profiles of DM halos. 

\subsection{Comparison to previous work}
\label{sSec_model_comp}

Studies aimed at providing theoretical predictions for the $c({\rm M},z)$ relation have traditionally
followed one of two routes. One class of models aims to connect the 
structural properties of halos to some aspect of their assembly history
\citep{Bullock2001,Eke2001,Wechsler2002,Zhao2003a,Lu2006,Maccio2008,Ludlow2014,Correa2015c}, 
most commonly by relating their characteristic
densities to the mean or background density at some appropriately
defined formation time. Other methods devise fitting formulae of
varying complexity that are then calibrated to the results of
numerical simulations over some range of halo mass and redshift 
\citep{Prada2012,Dutton2014,Diemer2015,Klypin2016}. 

Each of these methods have their own virtues and
weaknesses. Parametrized fits to simulation results, for example,
generally yield simple and compact formulae that accurately describe 
the scaling relations between halo structural properties, but must be 
treated with caution when extrapolated outside the range of halo mass,
redshift or cosmological parameters for which they were determined. 
Physically motivated models, however, attempt to link halo structure to some
aspect of their assembly. As a result they are often more cumbersome, 
but arguably provide more reliable extrapolations of these relations.  

In Fig.~\ref{fig:cmz_fits} we compare the predictions of our model 
(thin solid lines repeated in each panel) for
the Planck cosmology with several empirical fits proposed in the
literature (dashed lines). Different colours correspond to
different redshifts -- $z_0=0$, 1, 2, and 3 -- and all curves have 
been extrapolated down to $\sim 10^{-8} \, h^{-1}\, {\rm M}_\odot$ 
to emphasize their differences.

Each model was calibrated, at $z=0$, for masses 
$10^{10}\simlt {\rm {\rm M}_{200}}/[h^{-1}\, {\rm M}_\odot]\simlt 10^{15}$ 
and over that range provide consistent
predictions for the $c({\rm M})$ relation. Note, however, that the models 
show a larger variation at higher redshift, even over the mass 
scales at which they were calibrated (highlighted
using thick lines). This reflects the fact that
each group of authors used different equilibrium conditions (or none 
at all in the case of \citet{Diemer2015} and \citet{Prada2012}) to 
define their halo samples, which can introduce subtle biases in the 
recovered $c({\rm M},z)$ relation, particularly at high redshift 
\citep{Ludlow2012,Angel2016}. In addition, different authors
employ different techniques for estimating halo concentrations, which
may also account for some of the differences \citep[see, e.g.,][for a discussion]{Dutton2014}.

More importantly, however, these models fail to account for 
the effects of CDM free-streaming on halos of the smallest mass. 
This is shown in the lower-right panel, where the dashed lines
show the effect of a 100 GeV neutralino CDM particle relative to 
pure CDM (solid lines) on the predicted $c({\rm M},z)$ relation 
(we model the neutralino transfer function according to 
\citet{Green2004}). Note that the concentrations of neutralino 
halos peak at approximately 60 Earth masses, shown here
using a downward pointing arrow. Regardless of
the substructure mass function below this limit, the reduced
concentration of halos will substantially reduce the ``clumpiness''
of matter on scales comparable to Earth's mass and smaller. This
may have important implications for direct or indirect dark matter
detection experiments.

\section{Summary}
\label{SecConc}

We have analyzed the mass profiles and formation histories of a large
sample of equilibrium dark matter halos extracted from an ensemble of 
cosmological N-body simulations. These include the Copernicus Complexio
simulations, the Millennium and Aquarius simulations and an additional
set of runs with differing density fluctuation amplitudes and dark
matter content. Our simulation suite includes both CDM and 
WDM runs, allowing us to assess the dependence of
halo concentrations on cosmological parameters and on the shape of the
linear matter power spectrum. Below we summarize our main results. 

\begin{itemize}

\item In agreement with previous work, we find that the spherically
  averaged density profiles of CDM and WDM halos are approximately
  universal and are accurately described by the NFW or Einasto
  profile, regardless of the shape of the density fluctuation power
  spectrum. The main difference between CDM and truncated power
  spectra, such as WDM, is the mass-concentration-redshift
  relation of equilibrium halos.

\item The concentrations of CDM halos, at fixed $z$, decrease
  monotonically with increasing halo mass. At fixed mass,
  concentrations decrease monotonically with increasing redshift. We
  find no evidence for departures from this trend amongst our relaxed
  sample, providing further evidence that the recently reported
  ``upturn'' in the concentrations of rare, massive halos results from
  the inclusion of unrelaxed systems in the sample \citep[see,
  e.g.][for details]{Ludlow2012}. 

\item The WDM $c({\rm M})$ relation, on the other hand, is
  non-monotonic. At given $z$, concentrations peak at a multiple of
  the half-mode mass scale imposed by the power spectrum
  truncation. Concentrations decline above and below that peak mass
  scale (which seems to be independent of redshift) over the full mass
  range probed by the simulations.

\item The main-progenitor mass accretion histories (MAH) of CDM halos
  are scale-free, and resemble the NFW shape when cast in terms of
  mass and critical density, rather than mass and time. The
  concentration of a CDM halo can therefore be inferred from a simple
  scaling of its MAH. A simple model based on this feature accounts
  successfully for the main properties of the CDM
  mass-concentration-redshift relations.

\item The ``free-streaming'' truncation of the WDM linear power
  spectrum results in substantial delay and suppression of structure
  formation below a certain mass scale. This breaks the similarity of
  main-progenitor mass accretion histories, whose shape now depends
  critically on halo mass. Low-mass WDM halos near the truncation
  scale tend to form quickly and monolithically. Massive WDM halos, on
  the other hand, form just like their CDM counterparts. The strong
  dependence on halo mass of the MAH shape precludes the application
  of the same model developed to infer CDM halo concentrations from their
  MAHs.

\item A simple extension of the model, however, provides an excellent
  description of the mass-concentration-redshift relation for both CDM
  and WDM halos. The extension relies on using the full ``collapsed
  mass history'' of a halo (i.e., ${\rm M}_{\rm coll}(z)$, the total mass in
  collapsed structures at given time) rather than just its main
  progenitor. Collapsed mass fractions are approximately scale-free,
  which implies that a simple model may be devised in order to infer
  halo concentrations directly from ${\rm M}_{\rm coll}(z)$. This model
  reproduces well the mass-concentration-redshift relation in all our
  simulations, CDM and WDM alike, over the whole range of halo masses
  and redshifts probed. Applied to CDM, this model may be used to extrapolate $c({\rm
    M},z)$ to very low masses, where the relation departs strikingly
  from a pure power-law and differs also from the predictions of earlier models.

\end{itemize}

We provide upon request a simple code that computes the predicted $c({\rm M},z)$
relations for CDM and WDM power spectra with arbitrary free-streaming
truncation masses. This should be useful for extrapolating the results
we show here to halo masses and redshifts not well resolved by our
simulations. Such extensions may be important, especially when
evaluating predictions of these models for indirect and direct dark
matter detection strategies.

\section*{acknowledgments}

We thank Mark Lovell for useful conversations and our referee
for a constructive report that improved the paper.
ADL is supported by a COFUND Junior Research Fellowship;
SB by the STFC through grant ST/K501979/1.
LW acknowledges support from the NSFC grants program (No. 11573031,
No. 11133003), REA from AYA2015-66211-C2-2, and WAH from a Science 
and Technology Facilities Council grant ST/K00090/1
and the Polish National Science Center under contract 
\#UMO-2012/07/D/ST9/02785. 
This work was supported by the Science and Technology Facilities
Council (grant number ST/F001166/1) and European Research Council
(grant number GA 267291, ``Cosmiway'').
We also acknowledge financial support from the 
STFC consolidated grant ST/L00075X/1. This work used the COSMA Data 
Centric system at Durham University, operated by the Institute for 
Computational Cosmology on behalf of the STFC DiRAC HPC Facility 
(www.dirac.ac.uk). This equipment was funded by a BIS National 
E-infrastructure capital grant ST/K00042X/1, DiRAC Operations grant 
ST/K003267/1 and Durham University. DiRAC is part of the National 
E-Infrastructure.

\bsp
\label{lastpage}

\appendix
\section{Comparison to other physical $c({\rm M},z)$ models}
\label{AppComp}

\begin{figure*}
  \includegraphics[width=\textwidth]{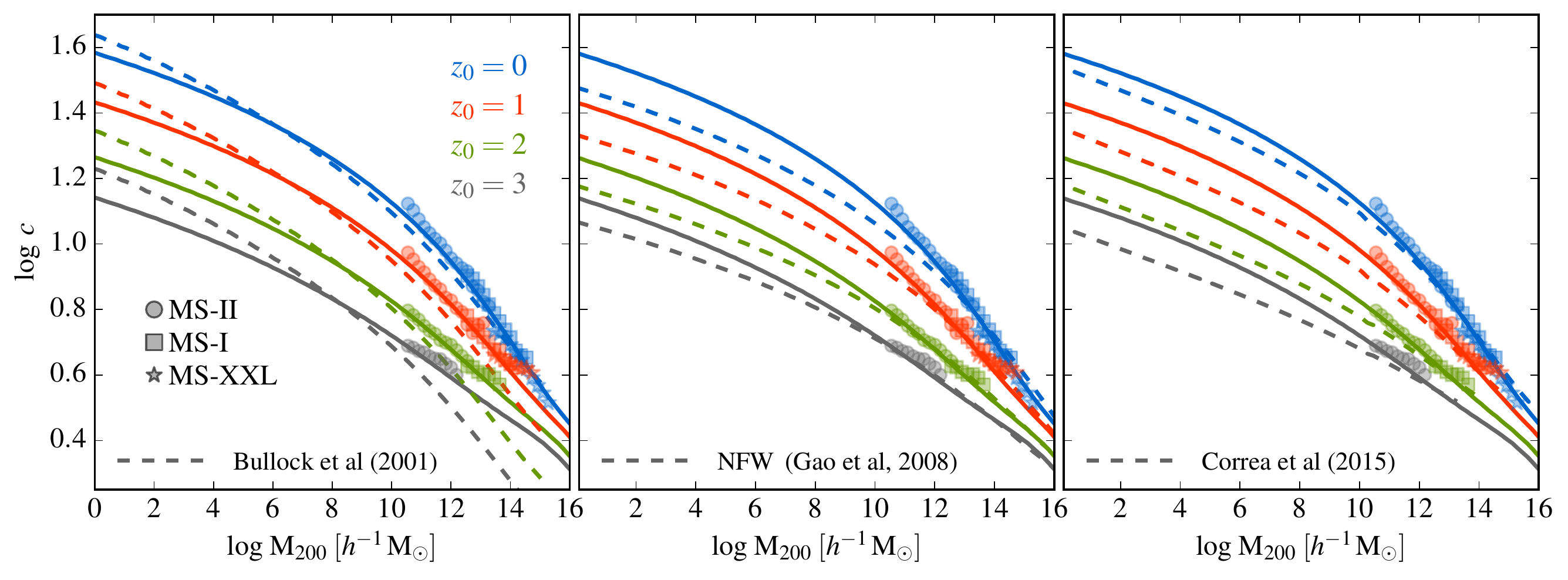}
  \caption{Comparison of the $c({\rm M},z)$ relations predicted by the
    analytic model described in Section~\ref{sSec_model} (shown using
    solid lines, and repeated in each panel) with several other proposals
    in the literature (dashed lines), including the
    \citet{Bullock2001} model, with modifications proposed by
    \citet{Maccio2008} (left); the NFW model, as modified by
    \citet{Gao2008} (middle); and the model of \citet{Correa2015c}
    (right).  For all models we have assumed a WMAP-1
    cosmology. Coloured symbols show the median $c({\rm M},z)$
    relations obtained for equilibrium halos in the Millennium
    simulations. Lines and symbols have been coloured according to
    redshift, as indicated in the legend.}
  \label{fig:cmz_models}
\end{figure*}

In Fig.~\ref{fig:cmz_models} we compare the predictions of the model
described in Section~\ref{sSec_model} to three common analytic recipes
to compute $c({\rm M},z)$ \citep{Bullock2001,Gao2008,Correa2015c} Each
model adopts a WMAP-1-normalized cosmology, and its predictions 
are compared to the median $c({\rm M},z)$ relations obtained for relaxed
halos in the MS runs.

All three analytic prescriptions considered here agree reasonably well
with our results at $z=0$. The \citet{Bullock2001} model, including
modifications suggested by \citet{Maccio2008} (left panel), deviates
by less than $\sim 5\%$ at any mass scale $\simlt 10^{15}\, h^{-1}\,
{\rm M}_{\odot}$.  This model predicts a simple redshift-dependence at fixed
mass, $c\propto (1+z)^{-1}$, which, at low masses ($\simlt 10^{10}\,
h^{-1} {\rm M}_{\odot}$), agrees well with our findings. At higher mass,
however, it predicts a sharp decline that is inconsistent with the
results of the Millennium simulations. 
For WDM halos this model predicts halo concentrations that increase
monotonically with decreasing mass, approaching a constant for masses
below the free-streaming scale (this is because Bullock et al define
formation times using $D(z_c)\sigma(F\, {\rm M}_0)=1.686$ which, for 
WDM, predicts a constant $z_c$ for ${\rm M}\ll {\rm M}_{\rm fs}$). 
Clearly this is not supported by the data in Fig.~\ref{fig:cmz}.

The modifications of the NFW model proposed by \citet{Gao2008} results
in concentration estimates that are similar, but not identical to the
predictions of the model we propose.  The original NFW model has three
free parameters: two physical parameters, $F$ and $f$, and a scaling
factor $C$ that relates the characteristic density of a halo to the
background density at its formation time. The formation time
corresponds to the time which a fraction $F$ of the halo's final mass,
${\rm M}_0$, was first assembled into progenitors each at least as
massive as $f\times {\rm M}_0$.

NFW chose $F=0.5$, $f=0.01$, and $C=3000$, whereas Gao et al found improved fits to the
Millennium simulation by choosing $F=0.1$, $f=0.01$ and $C=600$. 
Our proposal instead relates $F$ to the characteristic mass of the
halo, i.e. $F\equiv {\rm M}_{-2}/{\rm M_0}$; it adopts $f=0.02$ and $C=650$. We
compare the NFW predictions for these parameter choices 
in the middle panel of Fig.~\ref{fig:cmz_models}.

Note that defining $F\equiv {\rm M}_{-2}/{\rm M}_0$ reduces the number of model 
parameters while at the same
time improves the agreement with the simulation results. Note also,
that the Gao et al model is expected to fare increasingly poorly at mass
scales for which it predicts concentrations for which ${\rm
  M}_{-2}/{\rm M}_0\ll$ or $\gg$0.1. As a result, we
expect extrapolations of the Gao et al model to very low or very high mass
scales to become increasingly different. We provide a more detailed
discussion of the importance of the parameters $F$ and $f$ in Appendix
\ref{AppModel}.

The analytic predictions of \citet{Correa2015c}, shown 
in the right-hand panel of Fig.~\ref{fig:cmz_models}, also agree
well with our results. Like L14, this model relates
the characteristic density of a halo to the critical density at the 
redshift when its main progenitor had first assembled the
characteristic mass, ${\rm M}_{-2}$. The agreement is therefore not 
surprising, as their model is based 
explicitly on the results L14, which we deliberately
reproduce using our new methodology. Indeed, the slight differences
between the predictions of \citet{Correa2015c} and our own can be 
attributed to changes brought about by replacing Einasto fits with the NFW
fits upon which their model was calibrated. Note, however, that the
model of L14 (and Correa et al) fails when applied to WDM power
spectra because free-streaming modifies the proportionality constant
between the characteristic density of a halo and the critical density
at their particular choice of the collapse time. 

\section{Parameter-dependence of the characteristic density-formation time relation}
\label{AppModel}

\begin{figure*}
  \includegraphics[width=\textwidth]{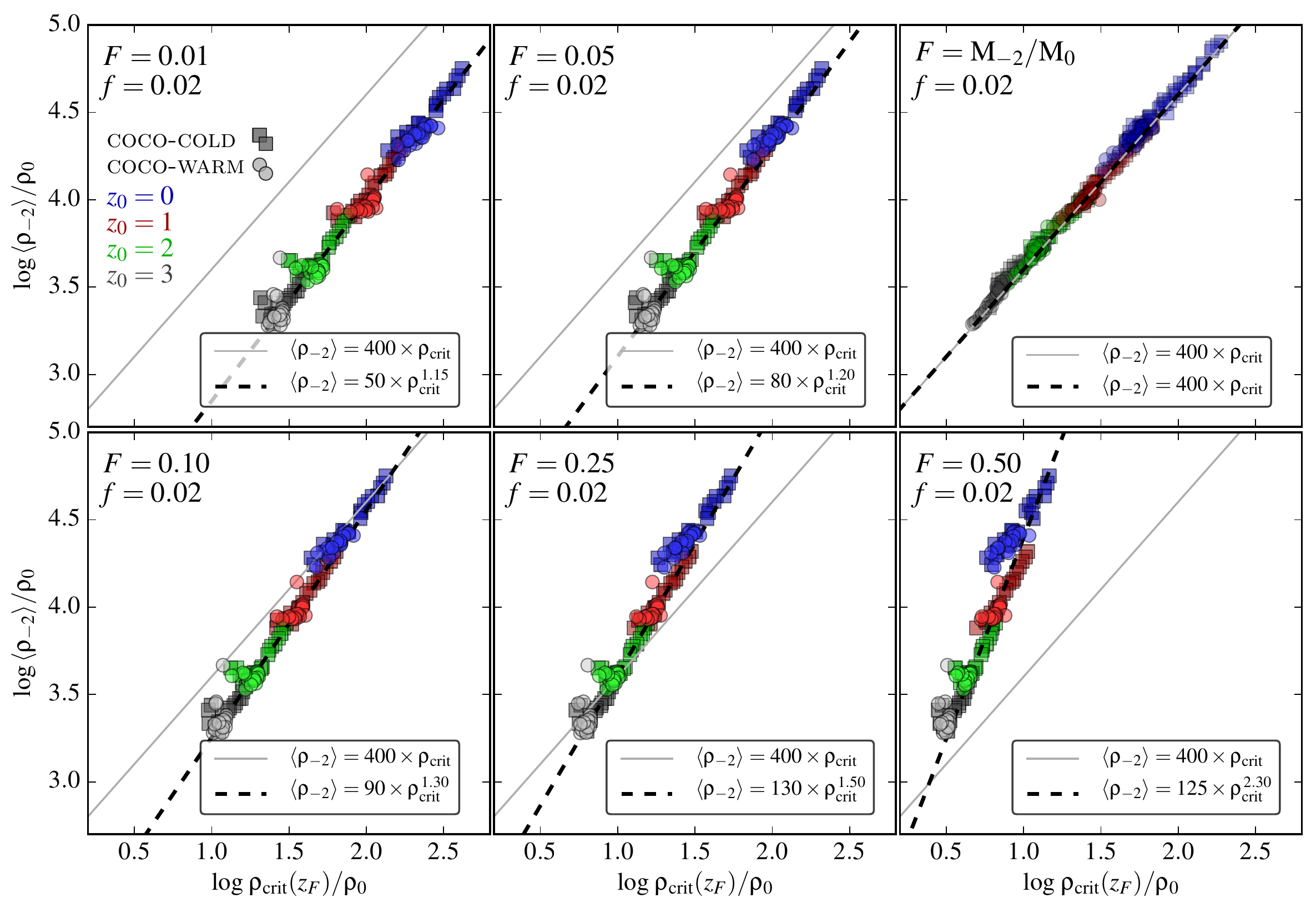}
  \caption{Mean enclosed density, 
    $\langle\rho_{-2}\rangle$, within the best-fitting scale radius,
    $r_{-2}$, versus the critical density at 
    the redshift $z_F$ at which a fraction $F$ of the halo mass, ${\rm M}_0$,
    was first contained in 
    progenitors more massive than $0.02\times {\rm M}_0$. Different panels
    correspond to different values of $F$. Enclosed and 
    critical densities have been scaled to $\rho_0=\rho_{\rm crit}(z_0)$, 
    the value at which each halo was identified. Different colours
    denote different identification redshifts, as indicated in 
    the legend. Point styles differentiate \textsc{coco-cold} (squares) from 
    \textsc{coco-warm} (circles), and correspond to the median relations
    obtained after averaging halo mass profiles and assembly histories
    in narrow bins of mass. Note that only halos with ${\rm N}_{200}\geq10^4$ 
    particles have been included. The thin grey line in each panel 
    shows the relation $\langle\rho_{-2}\rangle=400\times\rho_{\rm crit}(z_{-2})$, 
    which has the ideal one-to-one scaling, whereas the black dashed line shows a 
    power-law of different slope that was chosen to match the data.}
  \label{fig:rhorho_F}
\end{figure*}

\begin{figure*}
  \includegraphics[width=\textwidth]{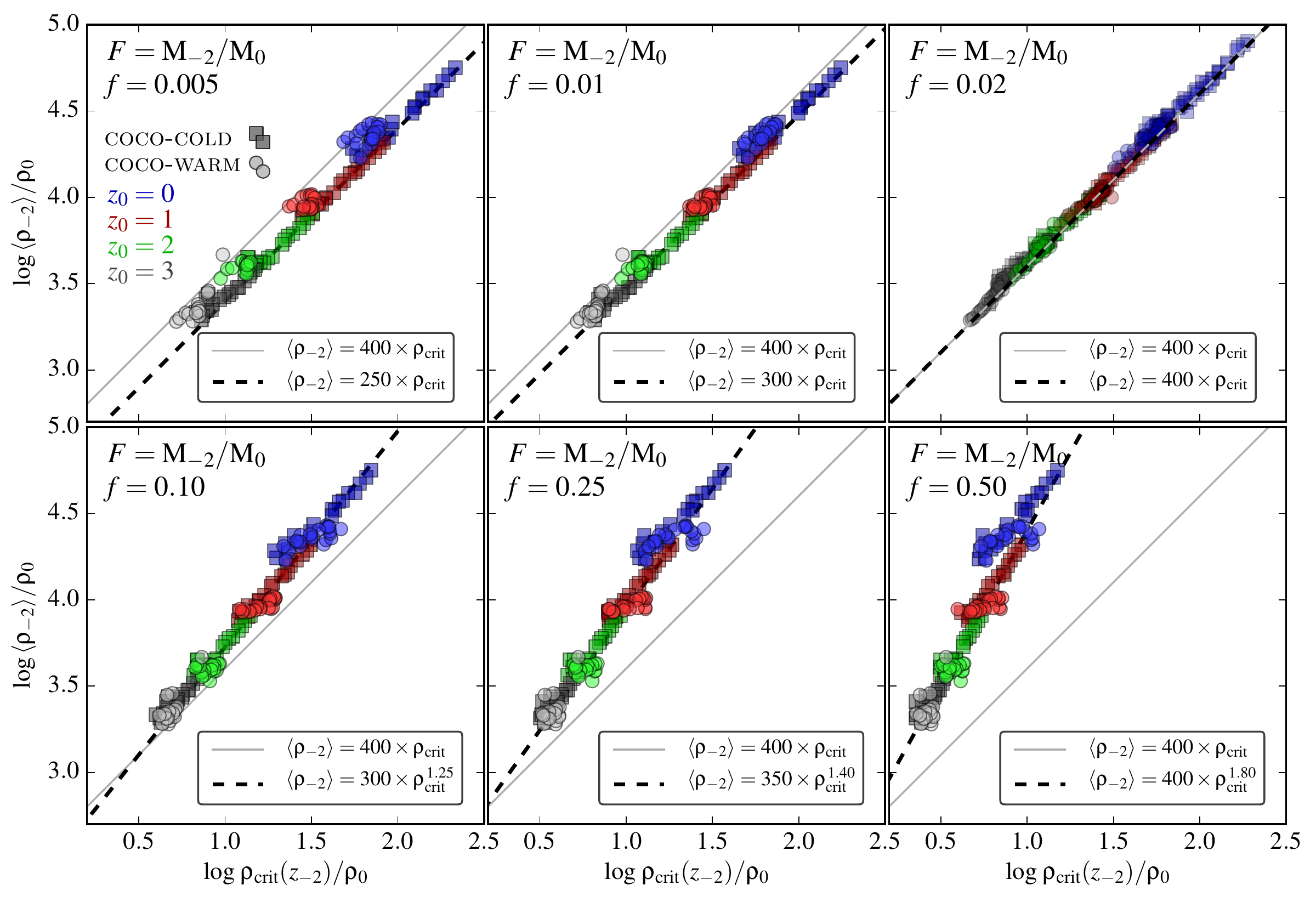}
  \caption{Same as Fig.~\ref{fig:rhorho_F}, but now showing how the 
    $\langle\rho_{-2}\rangle - \rho_{\rm crit}(z_{-2})$ scaling relation 
    depends on $f$, for $F={\rm M}_{-2}/{\rm M}_0$.}
  \label{fig:rhorho_f}
\end{figure*}

The model we propose here, like the NFW and Bullock
et al models, is afflicted by the introduction of two physical parameters,
$F$ and $f$, as well as a normalization constant, $C$, 
whose interpretation is unclear. The scaling
relations in Fig.~\ref{fig:rhorho} indicate that the value of $C$ will
depend sensitively on the precise definition of ``characteristic density''. If 
defined to be $\langle\rho_{-2}\rangle$, the mean density within
$r_{-2}$, then $C=400$. If, however, the characteristic density is
defined as that within $\sim 3\times r_{-2}$ then $C\approx 200$ describes
our numerical results quite well. Note that this is precisely what 
is expected from the spherical tophat collapse model. The simplest
interpretation of the parameter $C$ therefore is that it represents 
the contraction of the inner halo beyond what is expected from simple 
spherical collapse. Lacking a theoretical model to provide deeper 
insight, its numerical value must be calibrated using simulations,
and may exhibit subtle dependencies on the precise definition of halo mass
used. 

The success (of lack thereof) of previous attempts by NFW and Gao et al 
to model the $c({\rm M},z)$ relation using using methods similar to ours
can be traced to each authors different definition of ``formation time''. 
Both NFW and Gao et al assumed that a halo had ``formed'' when a fraction 
$F$ of its final mass had first assembled into clumps more massive than
a smaller fraction $f \ll F$ of the same mass. Each of these parameters
affects the precise value of $z_f$. NFW assumed $F=0.5$, Gao et al $F=0.1$,
and both found $f=0.01$ as a suitable progenitor mass threshold.

In Fig.~\ref{fig:rhorho_F}, we plot the mean enclosed density within
the halo scale radius, $\langle\rho_{-2}\rangle=M_{-2}/(4/3)\pi r_{-2}^3$, 
versus $\rho_{\rm crit}(z_{F})$, the critical density 
at the redshift $z_{F}$ when the total collapsed mass first
exceeded $F\times M_0$. Different panels
show results for different values of $F$, and all assume 
a small progenitor threshold, $f=0.02$. The points
corresponds to averages over single mass bins, and are shown for all four
identification redshifts after normalizing densities by the 
critical value at $z_0$, $\rho_{\rm crit}(z_0)$. Note that only mass
bins corresponding to halos with ${\rm N}_{200}\geq 10^4$ are plotted, which
ensures that all progenitors are resolved with at least 100 particles.

Although all values of $F$ (which span a factor of 50)
result in a tight linear relation between the two densities, 
only the particular choice $F= {\rm M}_{-2}/{\rm M}_0$ provides a direct proportionality between
$\langle\rho_{-2}\rangle$ and $\rho_{\rm crit}(z_{-2})$ (the thin grey
lines each each panel, for example, show the direct scaling
$\langle\rho_{-2}\rangle=400\times\rho_{\rm crit}(z_{-2})$, which
provides a reasonable description of the data only for $F= {\rm M}_{-2}/{\rm M}_0$).
All other values result in 
steeper slopes that gradually shallow as $F$ is decreased but , even 
for $F=0.01$, do not reach the natural linear scaling.

Formation times also depend on the progenitor mass threshold, $f$. Previous
studies have hinted at puzzlingly small values: $f=0.01$ in the 
case of NFW and Gao et al; $f=0.02$ in our case. In Fig.~\ref{fig:rhorho_f}
we show how the mean characteristic density, $\langle\rho_{-2}\rangle$,
of halos varies as a function of the redshift at which their characteristic mass,
${\rm M}_{-2}$, had first assembled into clumps each larger than $f\times {\rm M}_0$. 
Note that high values of $f$ (e.g. $f=0.5$, shown in the lower-right
panel) result in steep power law slopes. Decreasing $f$ shifts all
formation times to higher redshift, but the magnitude of the shift
depends on $\langle\rho_{-2}\rangle$. The net result,
provided $f$ becomes sufficiently small, provides a natural correspondence
between characteristic density and the background density at the halo 
formation time. Note also that, at least in the CDM case, the precise
values of $f$ seems unimportant, provided it is ``small enough'' (e.g.
$f=0.005$, 0.01 and 0.02 all result in scaling relations whose slopes
do not deviate noticeably from one). As a rule of thumb, we suspect
that values of $f\ll {\rm M}_{-2}/{\rm M}_0$, of order a few per cent,
will yield robust results.

Overall, we find that for $F={\rm M}_{-2}/{\rm M}_0$ and $f\approx 0.02$, the linear scaling, 
$\langle\rho_{-2}\rangle=C\times \rho_{\rm crit}(z_{-2})$, 
between these two densities is independent of both mass and
identification redshift. More importantly, however,
the zero-point of this relation is independent of the DM particle model:
both \textsc{coco-cold} and \textsc{coco-warm} have $C\approx 400$. 
We therefore advocate the use of these parameters in future modelling,
but acknowledge that better data, alternative halo definitions, or
or a different variety of DM models may result in modifications.

\begin{figure}
  \includegraphics[width=0.45\textwidth]{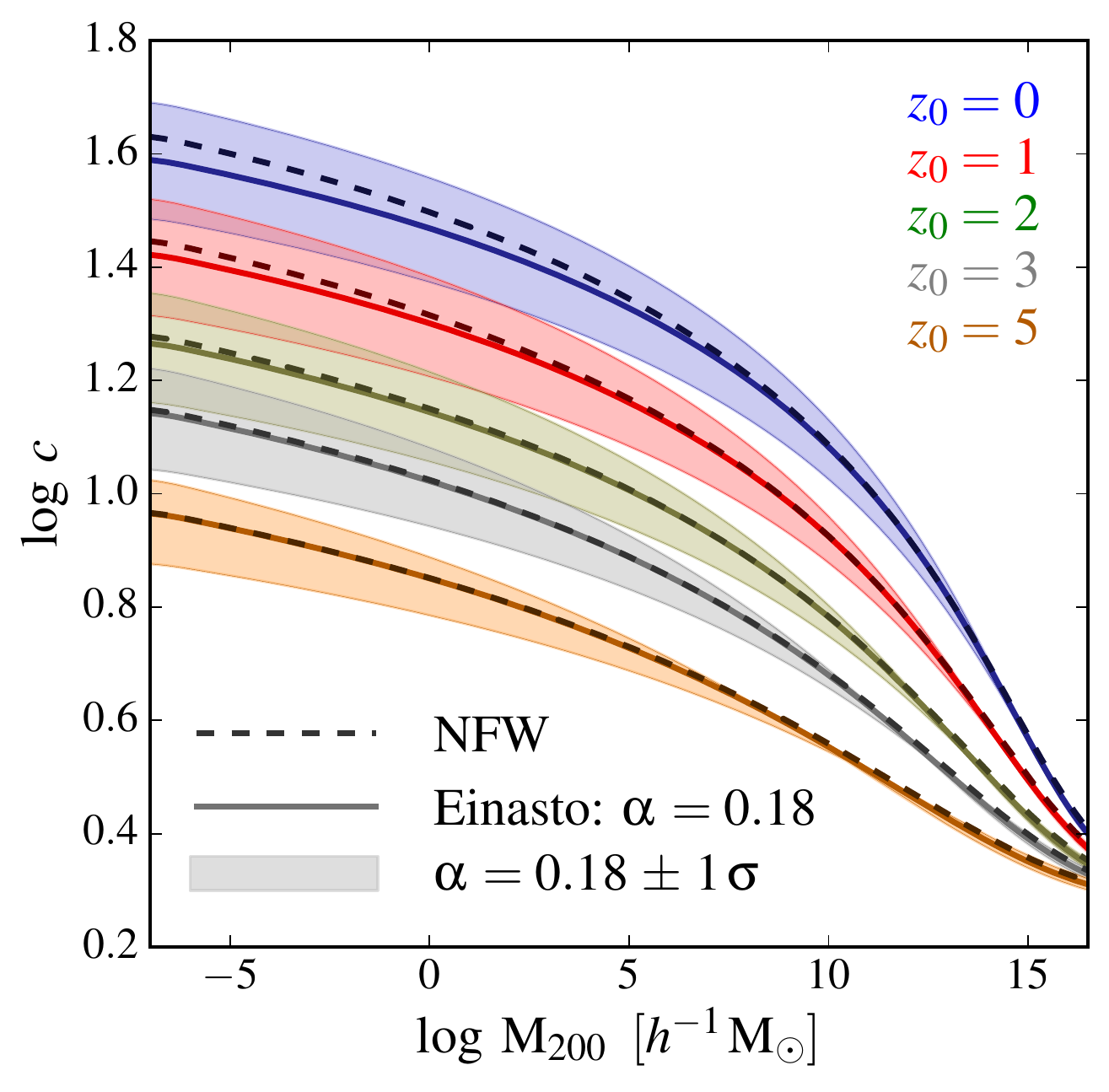}
  \caption{Predicted $c({\rm M},z)$ relations for different
    density profiles. Solid coloured lines assume an Einasto profiles
    with $\alpha=0.18$; dashed lines assume an NFW profile, and the
    shaded region highlights the range of $c$ expected given the 
    (redshift-dependent) scatter in $\alpha$ reported by
    \citet{Dutton2014}. Each model assumes $F={\rm M}_{-2}/{\rm M}_0$,
    $f=0.02$ and $C=650$. Note that at low-mass the intrinsic
    variation in the shapes of DM halo mass profiles may result in a
    substantial contribution to the scatter in concentration.}
  \label{fig:cmz_alpha}
\end{figure}

Mapping between characteristic densities and 
concentrations requires an assumption regarding the halo mass
distribution. Throughout the paper we have adopted an Einasto profile
with $\alpha=0.18$. In Fig.~\ref{fig:cmz_alpha} we plot the predicted
concentration-mass relations at several redshifts and for a few other
density profiles. The solid coloured lines correspond to our fiducial 
$\alpha=0.18$ Einasto profile, dashed lines to an NFW profile and the
shaded regions highlight the range of $c$ that is expected for
Einasto models with a redshift-dependent scatter in $\alpha$
consistent with the findings of \citet{Dutton2014}. Note that at high
mass, above $\sim 10^{10} \, h^{-1}\, {\rm M}_\odot$, the predicted
concentrations depend weakly on the assumed mass profile. Towards
lower mass, however, when concentrations become large, a modest 
halo-to-halo scatter in $\alpha$ can lead to a considerable scatter in the
predicted value of $c$.

\section{A fitting formula for the concentration-mass-redshift
  relation in the Planck cosmology}
\label{AppPlanck}

Our model for the $c({\rm M},z)$ relation, when expressed in terms of
dimensionless peak height, $\nu(z)=\delta_{\rm sc}/\sigma({\rm M},z)$, can be
accurately described by a broken power-law:
\begin{equation}
  c(\nu)=c_0\biggl(\frac{\nu}{\nu_0}\biggr)^{-\gamma_1}
  \biggl[1+\biggl(\frac{\nu}{\nu_0}\biggr)^{1/\beta}\biggr]^{-\beta (\gamma_2-\gamma_1)}.
  \label{cnu}
\end{equation}
Here $\delta_{\rm sc}=1.686$ is the spherical tophat collapse
threshold; $c_0$ and $\nu_0$ are normalizing constants; $\gamma_1$ and
$\gamma_2$ are the asymptotic power-law slopes towards low and high
$\nu$, respectively, and $\beta$ is the width of the transition
between these two regimes.

After some experimentation, we found that the values of these
parameters vary smoothly with redshift, $z$, and can be calibrated
once a cosmological model has been adopted. For the Planck cosmology,
their values may be reproduced as follows:
\begin{equation}
  c_0=3.395\times (1+z)^{-0.215},
  \label{A2}
\end{equation}
\begin{equation}
  \beta=0.307\times (1+z)^{0.540},
  \label{A3}
\end{equation}
\begin{equation}
  \gamma_1=0.628\times (1+z)^{-0.047},
  \label{A4}
\end{equation}
\begin{equation}
  \gamma_2=0.317\times (1+z)^{-0.893},
  \label{A5}
\end{equation}
and
\begin{equation}
  \begin{split}
    \nu_0=(4.135 - 0.564\, a^{-1} - 0.210\, a^{-2} \\
       +0.0557\, a^{-3} -0.00348\, a^{-4})\times  D(z)^{-1},
  \end{split}
\label{A6}
\end{equation}
where $a=(1+z)^{-1}$ and $D(z)$ is the linear growth factor. These
expressions are valid over the redshift range $1\geq\log(1+z)\geq 0$, 
and for masses $-8\leq \log {\rm M}/[h^{-1}\,{\rm M}_\odot]\leq 16.5$. 

\begin{figure*}
  \includegraphics[width=0.9\textwidth]{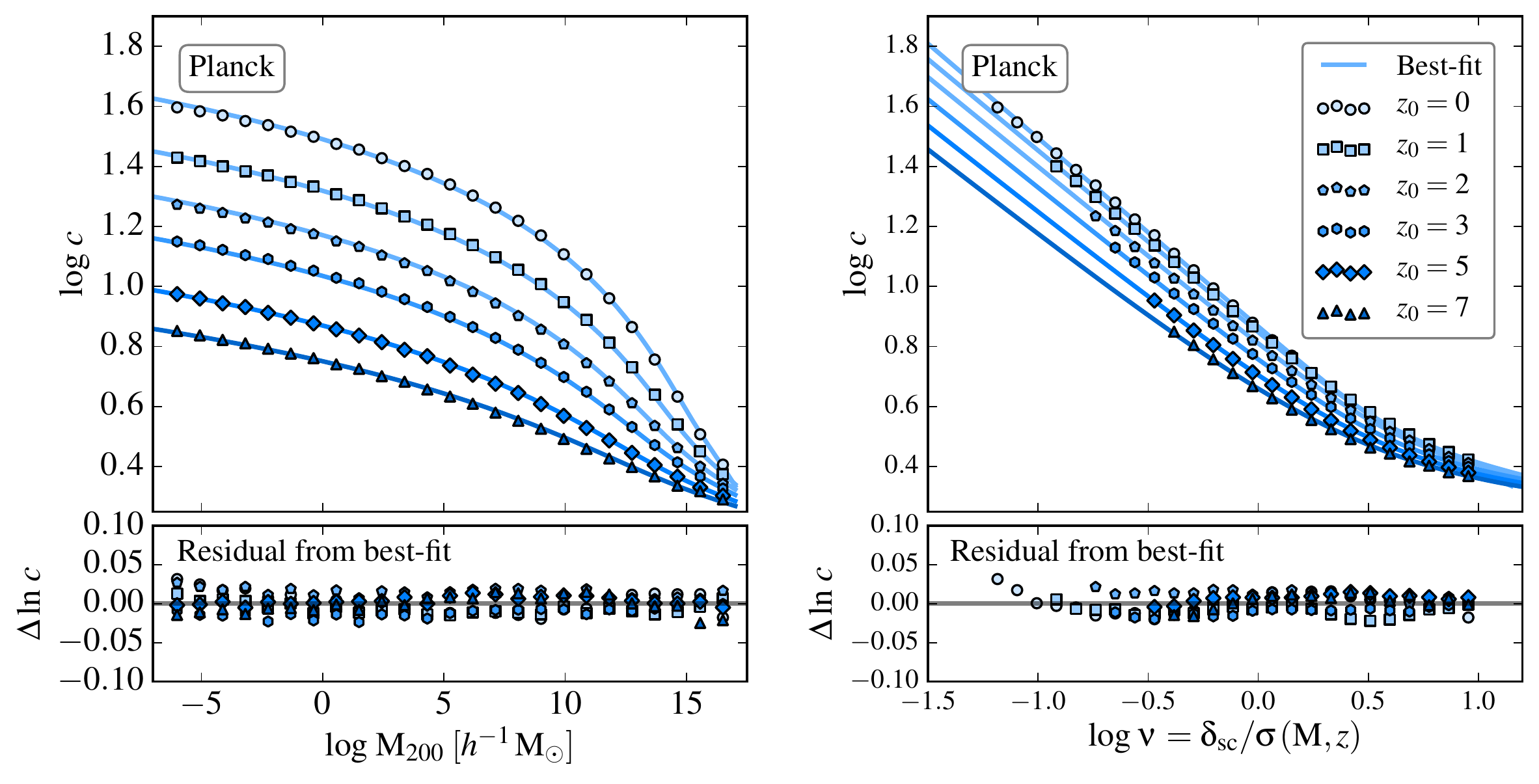}
  \caption{Dependence of halo concentration of mass (left panel) and
    peak height, $\nu$ (right panel), for several redshifts,
    $z_0$. Individual points show the predictions of the model
    described in Section~\ref{sSec_model} and dashed lines the
    best-fit relation constructed as described above.}
  \label{fig:cnu_planck}
\end{figure*}

In Fig.~\ref{fig:cnu_planck}
we compare the $c({\rm M},z)$ and $c(\nu)$ relations predicted by our model 
(coloured points) with the above fitting formula (solid lines). 
In agreement with previous studies
\citep[e.g.][]{Dutton2014,Diemer2015,Hellwing2016}, the $c(\nu)$ relation depends
slightly but systematically on redshift. Note also
that the residuals (shown in the lower panels) are small, typically less than
$\sim 3\%$ at {\em all} mass scales relevant for the CDM cosmology,
and show no systematic dependence on halo mass, ${\rm M}$, or
redshift.

Mapping between peak height and halo mass is achieved via the rms
density fluctuations, defined
\begin{equation}
  \sigma({\rm M},z)=\frac{1}{2\pi^2}\int^{\infty}_0 P(k,z) W^2(k,{\rm M}) k^2 \, dk.
  \label{rms}
\end{equation}
Here $P(k,z)$ is the linear fluctuation power spectrum as
a function of wavenumber $k$, and $W(k,{\rm M})$ is the Fourier transform of
the spherical top-hat window function. For halo masses spanning
$10^{-7}\simlt {\rm M}/[h^{-1}\, {\rm M}_\odot] \simlt 10^{15}$,
$\sigma({\rm M},z)$ can be approximated to
within $\sim 2.5\%$ by
\begin{equation}
  \sigma({\rm M},z)=D(z)\frac{22.26\, \xi^{0.292}}{1+1.53\,
    \xi^{0.275}+3.36\, \xi^{0.198}},
  \label{rms_fit}
\end{equation}
where 
\begin{equation}
  \xi\equiv\biggr(\frac{{\rm M}}{10^{10} \, h^{-1}\, M_\odot}\biggl)^{-1}.
  \label{Mass}
\end{equation}
The linear growth factor can be approximated by \citep{Lahav1991}
\begin{equation}
  D(z)=\frac{\Omega_{\rm M}(z)}{\Omega_{\rm M}^0}\frac{\Psi(0)}{\Psi(z)}(1+z)^{-1},
  \label{GF}
\end{equation}
with
\begin{equation}
\Psi(z)=\Omega_{\rm M}(z)^{4/7}-\Omega_{\Lambda}(z)+\biggl(1+\frac{\Omega_{\rm M}(z)}{2}\biggr)\biggl(1+\frac{\Omega_{\Lambda}(z)}{70}\biggr),
\end{equation}
\begin{equation}
\Omega_{\Lambda}(z)= \frac{\Omega_{\Lambda}^0}{\Omega_\Lambda^0 + \Omega_{\rm M}^0 (1+z)^3},
\end{equation}
and $\Omega_{\rm M}(z)=1-\Omega_{\Lambda}(z)$. Note that $\Omega_i^0$
refers to the {\em present-day} contribution to the critical density
from component $i$.

In the Planck cosmology, the concentration of any halo of mass ${\rm M}$ at redshift $z$ 
can therefore be estimated as follows:

\begin{enumerate}
\item Calculate $\nu$ using eqs~\ref{rms_fit}, \ref{Mass} and
  \ref{GF}.
\item Evaluate eqs~\ref{A2} through \ref{A6}, for redshift $z$. 
\item Use the resulting values of these parameters to calculate $c$ from eq.~\ref{cnu}.
\end{enumerate}

\bibliographystyle{mn2e}
\bibliography{paper}

\end{document}